\documentclass[pra,twocolumn,superscriptaddress,amsmath,amsfonts]{revtex4-2}

\usepackage{graphicx,float}
\usepackage{braket}
\usepackage{siunitx}
\usepackage{appendix}
\usepackage{amssymb}
\usepackage{bm}
\usepackage{mathrsfs,mathrsfs}
\usepackage{slashed}
\usepackage[colorlinks]{hyperref}
\usepackage{xcolor}

\begin{document}

\title{Using system-reservoir methods to derive effective field theories for broadband nonlinear quantum optics: a case study on cascaded quadratic nonlinearities}
\author{Chris Gustin}
\thanks{These authors contributed equally to this work; email: cgustin@stanford.edu}
\affiliation{E.\,L.\,Ginzton Laboratory, Stanford University, Stanford, California 94305, USA}
\author{Ryotatsu Yanagimoto}
\thanks{These authors contributed equally to this work; email: cgustin@stanford.edu}
\affiliation{Physics \& Informatics Laboratories, NTT Research, Inc., Sunnyvale, California 94085, USA}
\affiliation{School of Applied and Engineering Physics, Cornell University, Ithaca, New York 14853, USA}
\affiliation{E.\,L.\,Ginzton Laboratory, Stanford University, Stanford, California 94305, USA}
\author{Edwin Ng}
\affiliation{Physics \& Informatics Laboratories, NTT Research, Inc., Sunnyvale, California 94085, USA}
\affiliation{E.\,L.\,Ginzton Laboratory, Stanford University, Stanford, California 94305, USA}
\author{Tatsuhiro Onodera}
\affiliation{School of Applied and Engineering Physics, Cornell University, Ithaca, New York 14853, USA}
\affiliation{Physics \& Informatics Laboratories, NTT Research, Inc., Sunnyvale, California 94085, USA}
\affiliation{E.\,L.\,Ginzton Laboratory, Stanford University, Stanford, California 94305, USA}

\author{Hideo Mabuchi}
\affiliation{E.\,L.\,Ginzton Laboratory, Stanford University, Stanford, California 94305, USA}

\date{\today}
 \begin{abstract}
 In broadband quantum optical systems, nonlinear interactions among a large number of frequency components induce complex dynamics that may defy heuristic analysis. In this work we introduce a perturbative framework for factoring out reservoir degrees of freedom and establishing a concise effective model (effective field theory) for the remaining system. Our approach combines approximate diagonalization of judiciously partitioned subsystems with master equation techniques. We consider cascaded optical $\chi^{(2)}$ (quadratic) nonlinearities as an example and show that the dynamics can be construed (to leading order) as self-phase modulations of dressed fundamental modes plus cross-phase modulations of dressed fundamental and second-harmonic modes. We then formally eliminate the second-harmonic degrees of freedom and identify emergent features of the fundamental wave dynamics, such as two-photon loss channels, and examine conditions for accuracy of the reduced model in dispersive and dissipative parameter regimes. Our results highlight the utility of system-reservoir methods for deriving accurate, intuitive reduced models for complex dynamics in broadband nonlinear quantum photonics.
\end{abstract}
\maketitle
\section{Introduction}

Since the first demonstrations of fundamental concepts in quantum mechanics with photons~\cite{Bouwmeester1997,Furusawa1998,Pan1998,Aspect1982}, nonlinear optics has played central roles in manipulating the quantum nature of light. Beyond explorations of fundamental science, such an ability holds promise to unlock functions beyond classical limitations in a wide range of applications~\cite{LIGO2013, Tsang2016, Duan2001}. To this goal, solid state nonlinear optics offers a highly scalable and room-temperature approach, compared to more traditional quantum platforms, e.g., cavity and microwave quantum electrodynamics. While the lack of strong photon-photon interactions in waveguides with bulk optical nonlinearities has so far has only allowed for experimental access to Gaussian quantum states~\cite{Braunstein2005} (i.e., coherent and squeezed states), rapid progress in the emergent field of broadband nonlinear nanophotonics~\cite{Hickstein2019Jul,Jankowski2021Sep,Mishra2022Aug,Singh2020,Nehra2022few,Solntsev2022Jun} enables tight three-dimensional field confinements on a low-loss platform with advanced dispersion-engineering, which has significantly bridged the gap towards the strong coupling regime~\cite{Yanagimoto2022Nov}. Consequently, it is likely that nonlinear optics will enter a qualitatively distinctive regime beyond the traditional Gaussian formalism in the near future, where highly complicated \emph{non-Gaussian} quantum dynamics take place among multiple frequency modes. 

Such dynamical complexity is both a blessing and a curse: non-Gaussian quantum features~\cite{Walschaers2021} let us explore the entire range of quantum states of light, e.g., for quantum information processing~\cite{Langford2011Oct,Takeda2019}, and a multimode implementation could offer unique potential for quantum simulation and light sources. At the same time, such quantum dynamics can, in principle, be exponentially complicated and computationally intractable without simplifying assumptions~\cite{Lloyd1999}. Therefore, to fully leverage the experimental capabilities in broadband nonlinear nanophotonics that are expected to be available in the near future, it is essential to advance understanding in various phenomenologies and establish sophisticated \emph{model reduction} frameworks in which to concisely and rigorously understand each of them.

In quantum optics, arguably the most effective tool for model reduction is the notion of an open quantum system~\cite{breuer2002,Gardiner}. In an optical cavity, for instance, resonant photons remain localized within the cavity only temporarily, before scattering into the environment, and a full model of such a process (including the entire environment) is neither tractable nor desirable. A notion of a ``system'' and ``reservoir'' is thus introduced, and the influence of the larger reservoir on the system is incorporated using techniques of open quantum systems---for example, the master equation (ME). 
The result is a concise effective model of only the system coordinates' dynamics.

Often, the separation between the system and reservoir is simple enough that a simple phenomenological approach suffices to capture the features the reservoir imparts on the system. However, there is a growing recognition of the need to develop \emph{ab-initio} models of such a system-reservoir construction~\cite{Lentrodt2020Jan,Salmon2022Mar,Franke2019May,Lentrodt2023Jun,Onodera2022Mar}, driven by advances in light-matter interaction strengths in a variety of platforms~\cite{FriskKockum2019Jan}. Moreover, in a generic many-body system, there exist infinite possible ways to draw system-reservoir boundaries within a system. Therefore, it is tempting to ask to what extent this notion of system-reservoir partition can be extended to concisely formulate intra-system many-body quantum dynamics.

In this work, we show that such methods can be a highly powerful framework in which to understand certain \emph{closed}-system dynamics in broadband nonlinear photonics. Specifically, as a case study, we consider a paradigmatic example of broadband cascaded $\chi^{(2)}$ interactions~\cite{Xu2002, Kobyakov1996, Menyuk1994, DeSalvo1992}, where fundamental harmonic (FH) and second harmonic (SH) modes interact via second-order nonlinear interactions with a large center-band phase-mismatch. In classical nonlinear optics, the cascaded $\chi^{(2)}$ process is often employed as a means to realize effective $\chi^{(3)}$ interactions of the FH modes, for which we provide a concise explanation through the lens of quantum system-reservoir-like methods. We show that one can perceive the FH and SH modes as ``system'' and ``reservoir'' subsystems, respectively, and derive effective models that reproduce the effective $\chi^{(3)}$ interactions of the FH modes, and predict additional higher-order features. Equivalently, this procedure can be seen as constructing an effective field theory (EFT) for the relevant system degrees of freedom.
Interestingly, different system-reservoir boundaries can be more natural, depending on the dispersion relationships among the modes involved. 
While we focus on the example of cascaded $\chi^{(2)}$, our results are general and could be applied to other multimode photonic systems, e.g., with $\chi^{(3)}$ interactions, or more generally to other 1D interacting many-body bosonic systems. We expect our work to highlight the ubiquity of system-reservoir methods, showing that they can concisely explain multimode quantum phenomena in broadband nonlinear photonics.

Below we describe the structure of the article and our presented framework in detail with a more technical summary of each section.

\subsection*{Introduction to the EFT framework} In Sec.~\ref{sec:general}, we outline our approach to deriving effective field theories (EFTs) for broadband nonlinear optics in general terms. These concisely capture the essential dynamical features by leveraging emergent system-reservoir partitions within a multimode quantum system, while conveying extra analytic insights and retaining an ab-initio approach. As shown in Fig.~\ref{fig:schematic0} (a), we assume the system to have multiple frequency bands that are perturbatively coupled to each other via natural nonlinear optical interactions, with the local band energy structure determined by underlying dispersion relations. We present techniques to eliminate the ``reservoir'' bands in favor of a reduced model for the system bands.
Our method is not limited to quantum optics, and can be applied generally to any bosonic many-body quantum system that can be described as weakly-interacting energy bands. The general approach is to use a Schrieffer-Wolff (SW)~\cite{Schrieffer1966, Bravyi2011, LeBoite2020Jul}  transformation to eliminate dispersive couplings to phase-mismatched bands, where residual couplings between the two coupled systems can then be addressed by using a mean-field approximation~\cite{Degenfeld-Schonburg2014Jun}, wherein the dynamics of the band to be eliminated are simply replaced with their free evolution solutions. Dissipative coupling channels are treated by tracing out the band in which photons decay into by means of a Born-Markov ME, which can be applied to dispersive interactions as well. The Markovianity of such an effect is determined by the group velocity mismatch of the photons involved in the exchange. Our approach is highly generalizable and is compatible with many of the standard tools of analysis in quantum optics and quantum engineering.

\subsection*{Case study: cascaded quadratic nonlinearities} Following the general approach outlined in Sec.~\ref{sec:general}, we use the example of cascaded quadratic optical nonlinearities as a case study for our approach, and analyze this case in detail. 
While cascaded quadratic nonlinearities have long been studied and used as a means to realize strong cubic nonlinearities~\cite{Xu2002, Kobyakov1996, Menyuk1994, DeSalvo1992} benefiting from the large magnitude of $\chi^{(2)}$ nonlinear coefficients~\cite{Boyd2008}, their theoretical treatments have often been based on classical coupled wave equations, and it is vital to establish a rigorous quantum mechanical theory for future quantum applications. This has been done recently using a ME approach for resonant cavity modes~\cite{Onodera2022Mar}. Here we extend this result to the case of a broadband multimode continuum.

Depending on the underlying dispersion relations, this system admits two regimes we study: a ``dispersive coupling'' regime, where the dynamics of the dressed FH modes reduces to that of cubic nonlinearity, and a ``dissipative coupling'' regime, where the cubic nonlinearity is supplemented by a small effective two-photon loss channel to red and blue detuned dispersive waves, arising from an elliptical dissipation surface solution to the phase-matching condition.

Additionally, we show that dressed FH and SH modes interact via cross-phase modulations without exchanging excitations, which provides a flexible means to realize an effective spatial potential for the dressed FH photons. Such a spatial potential can play essential roles in various many-body effects~\cite{Yao2020, Sutherland1971, Pitaevskii1961}, highlighting the potential of cascaded quadratic nonlinearities for quantum simulations~\cite{Aspuru2012, Noh2016} or computation~\cite{Yanagimoto2022Nov}.

 After presenting in Sec.~\ref{sec:model} a fully quantum mechanical model of propagating FH and SH multimode fields in a $\chi^{(2)}$-nonlinear waveguide, in Sec.~\ref{sec:classical}, we sketch a heuristic classical mechanical derivation of the effective cubic cascaded nonlinearity for multimode fields, which corresponds to self-phase modulation (SPM) in the spatial basis. In Sec.~\ref{sec:sw}, we use the SW transformation and Born-Markov ME to derive effective models with cubic nonlinearities and higher-order effects, and verify the convergence with the full $\chi^{(2)}$ Hamiltonian in the large phase-mismatched limit with numerical simulations. In Sec.~\ref{sec:mf} we address the issue of the transformation between lab and SW frames and its effect on observables and quantum dynamics, particularly with regard to higher-order effects. We present a semi-analytic mean-field theory to account for the non-zero SH amplitudes in the SW frame, and also show how this subtlety can be circumvented by the use of an adiabatic modulation of the phase-mismatch. Sec.~\ref{sec:conclusions} presents conclusions.

We also include six appendices. In Appendix~\ref{app:nondimensionalization} we give further details on the nondimensionaliation of the multimode $\chi^{(2)}$ Hamiltonian. In Appendix~\ref{app:higherorder}, we construct the next-to-leading order expansion term in the SW expansion, and use it to demonstrate the order of validity to which our perturbative Hamiltonian is valid. In Appendix~\ref{app:mesons}, we shed light on the physical interpretation of the dressed SH excitations, revealing them to be the so-called ``optical mesons''~\cite{Drummond1997}. Appendix~\ref{app:master_equation} presents a ME model of the cascaded nonlinearity, by tracing out the SH as a reservoir, and we compare and contrast the (purely) ME method with the SW approach. In Appendix~\ref{app:diss_me}, we provide a derivation of the ME used to eliminate the extra-band component of the FH in the main text, and we also show how the phase-mismatch between the center of FH and SH bands determines the memory kernel of the interaction, and how sufficiently large phase-mismatch leads to a Markovian decay regime in which the Born-Markov ME is valid. Finally, in Appendix~\ref{app:single}, we consider a toy model of a single-moded (in both the FH and SH) system which retains the features of the dissipative coupling regime, and use this model to numerically demonstrate how adiabatic modulation of the quasi-phase-matching poling period can simplify the modelling of the dressed-frame dynamics by physically initializing the system in the eigenstates of its constituent subsystems.

\section{Outline of general method of band elimination}\label{sec:general}
\begin{figure*}
\includegraphics[width=0.8\textwidth]{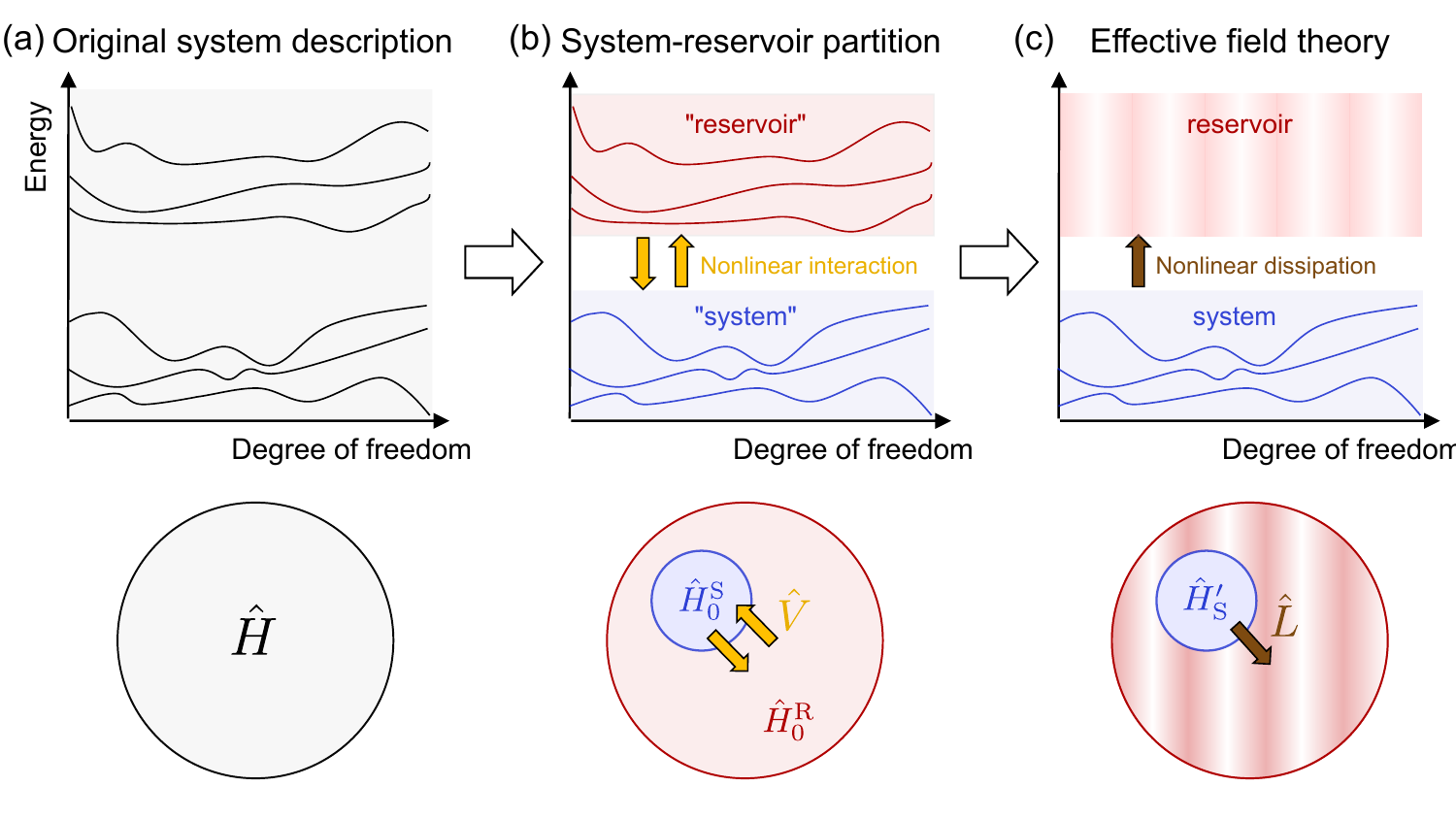}
    \caption{Schematic of EFT method for multi-band quantum systems. (a) In original system description, we perceive the whole multimode quantum system as a single system with Hamiltonian $\hat{H}$. (b) To concisely capture the physics, we partition the system into the ``system'' part and ``reservoir'' part where their free-evolutions are described by $\hat{H}_0^\mathrm{S}$ and $\hat{H}_0^\mathrm{R}$, respectively. The interaction between these parts are described by $\hat{V}$. At this point, the partition is only conceptual (and thus quotation marks on ``system'' and ``reservoir''). (c) Based on the system-reservoir partition, we derive an effective field theory, where the coupling of the system to the reservoir is reduced to nonlinear dissipation, represented by potential Lindblad operator(s) $\hat{L}$. The internal dynamics of the system is modified to the form $\hat{H}'_\mathrm{S}$.
    }
    \label{fig:schematic0}
\end{figure*}

In this section, we sketch some of the main ideas of this work on how to obtain effective reduced models for multimode nonlinear quantum optics in the presence of perturbative band-band couplings. We refer to excitation quanta as photons and energy bands as dispersion relations, but the procedure is general to bosons in 1D. Here we only highlight the most important features---for a more detailed exposition, we refer the reader to Sec.~\ref{sec:model} and onward, where we focus on the case of cascaded quadratic optical nonlinearities. Figure~\ref{fig:schematic0} gives a conceptual schematic of the approach.

In general, we shall assume the system to be described by a Hamiltonian of the form $\hat{H} = \hat{H}_0 + \hat{V}$, with $\hat{H}_0 = \sum_i \hat{H}_0^{(i)}$, where
\begin{equation}
    \hat{H}_0^{(i)} =  \int_{\mathcal{P}_i}dk E_i(k) \hat{\phi}^{\dagger}_i(k)\hat{\phi}_i(k)
\end{equation}
represents the bare energy of the $i$th band. Here, we assume that these correspond to a partitioning of the 1D electromagnetic fields into disjoint frequency regions with energy (dispersion) band $E(k)$, though they could also more generally correspond to entirely different physical fields, e.g., polarization or spatial modes. Formally, we can consider the entire Hilbert space $\mathcal{H} = \bigotimes_{i}\mathcal{H}_i$, where $\mathcal{H}_i$ corresponds to the $i^{\rm th}$ photon band with creation and destruction operators which satisfy $[\hat{\phi}_i(k),\hat{\phi}_j^{\dagger}(k')]=\delta_{ij}\delta(k-k')$. The frequency partitions are centered about $k = \bar{k}_i$ such that $\mathcal{P}_i = [\bar{k}_i - \Delta_i^{(-)}, \bar{k}_i + \Delta_i^{(+)}]$, where  $\Delta_i^{(+)} - \Delta_i^{(-)}$ gives the $i^{\rm th}$ partition bandwidth.

In the context of this work, instead of perceiving the entire Hilbert space as a single closed system, we consider a partition $\mathcal{H} = \mathcal{H}_{\rm S} \otimes \mathcal{H}_{\rm R}$, where $\mathcal{H}_\mathrm{S}$ and $\mathcal{H}_\mathrm{R}$ denotes ``system'' and ``reservoir'' subsystems, respectively. The exact decomposition will depend on the specifics of the system under study. As shown in later sections, an appropriately chosen boundary leads to a concise reduced model under only $\mathcal{H}_{\rm S}$, which provides unique insights that are hard to obtain in the original model. 

Crucially, the ``bare'' Hamiltonian $\hat{H}_0$ acts on each subsystem individually, and does not generate any entanglement between them. Nontrivial dynamics occur when couplings among the subsystems are considered, for which the interaction Hamiltonian is given generically as 
\begin{equation}
    \hat{V} = g_0\int d\mathbf{k} G(\mathbf{k})\hat{S}^{\dagger}_{\mathbf{k}}\hat{R}_{\mathbf{k}} + \text{H.c.},
\end{equation}
where here the vector $\mathbf{k} = (k_1, k_2, ..., k_N)$ denotes not a Euclidean space wavevector, but rather a tuple of reciprocal-space integration variables. In the nonlinear optics case,
$N+1$ denotes the number of photons involved in the interaction. For instance, $N=1$ corresponds to linear coupling between two bosonic modes with
\begin{align}
    \hat{V}=g_0\int dk_1G(k_1)\hat{\phi}_1^\dagger(k_1)\hat{\phi}_2(k_1).
\end{align}
$\hat{S}_{\mathbf{k}}$ and $\hat{R}_{\mathbf{k}}$ are operators of the $\mathcal{H}_{\rm S}$ and $\mathcal{H}_{\rm R}$ subsystems, respectively, which we assume to be normally-ordered products of annihilation and creation operators. For the case of linear coupling above, these operators took the form $\hat{S}_{k_1} = \hat{\phi}_1(k_1)$ and $\hat{R}_{k_1} = \hat{\phi}_2(k_1)$. The generalization to the case in which there are multiple system and reservoir operators constituting $\hat{V}$ is straightforward. 

Moving forward, we shall assume the dimensionless coupling function $G(\mathbf{k})=1$ for simplicity. Generalizing the results in this work to arbitrary $G(\mathbf{k})$ is straightforward, but may affect the validity of the approximations made---particularly the Born-Markov approximation. Generally, as long as $G(\mathbf{k})$ changes on a characteristic scale slower than all the other parameters of the system Hamiltonian, all results should remain valid~\cite{Onodera2022Mar}. 

As a consequence of the assumptions made,
\begin{subequations}\label{eq:S_comm}
    \begin{equation}[\hat{S}_{\mathbf{k}},\hat{H}_0^{\rm S}] = \mathcal{E}_{\rm S}(\mathbf{k})\hat{S}_{\mathbf{k}},
    \end{equation}
        \begin{equation}
        [\hat{R}_{\mathbf{k}},\hat{H}_0^{\rm R}] = \mathcal{E}_{\rm R}(\mathbf{k})\hat{R}_{\mathbf{k}},
    \end{equation}
    \end{subequations}
    where $\mathcal{E}_{\rm S}(\mathbf{k})$ ($\mathcal{E}_{\rm R}(\mathbf{k})$) is the \emph{net} bare photon energy removed by the operator $\hat{S}_{\mathbf{k}}$ ($\hat{R}_{\mathbf{k}}$), determined by the appropriate summation over the single-photon bare energy bands $E_i(k)$. 

For a given set of $\hat{H}_0$ and $\hat{V}$, we can derive reduced models for the dynamics within the system subsystem $\mathcal{H}_\mathrm{S}$ using several techniques, depending on the nature of the interactions. Below, we introduce two main techniques that we employ in this work: SW transformation and Born-Markov ME.

\subsection{Dispersive interactions --- Schrieffer-Wolff approach}

For a system with Hamiltonian $\hat{H}=\hat{H}_0+\hat{V}$, the SW formalism aims at identifying a near-unit unitary transformation $\hat{U}$ that eliminates the perturbative coupling term $\hat{V}$ in the interaction frame. The SW transformation was first introduced to establish equivalence between the Kondo model and Anderson impurity model~\cite{Schrieffer1966}; here, we employ the formalism as a central theoretical tool to analyze emergent system-reservoir physics with off-resonant (dispersive) couplings.

The SW transformation $\hat{U}=e^{\hat{A}}$ is characterized by an anti-Hermitian operator $\hat{A}$ that satisfies
 \begin{equation}\label{eq:comm_req}
     [\hat{A},\hat{H}_0] = -\hat{V}.
 \end{equation}
 If such $\hat{A}$ is found, the resultant unitary transforms the Hamiltonian via the Baker-Campbell-Hausdorff formula as 
 \begin{equation}\label{eq:SW_simpl}
     \hat{U}\hat{H}\hat{U}^{\dagger} = \hat{H} + \sum_{n=1}^{\infty} \frac{n}{(n+1)!}\left[\mathcal{C}(\hat{A})\right]^n[\hat{V}],
 \end{equation}
 where $\mathcal{C}(\hat{A})[\hat{o}] = [\hat{A},\hat{o}]$ is the commutator superoperator. Clearly, if $\hat{A} \sim \epsilon \ll 1$ for some small parameter $\epsilon$, we can construct a perturbative expansion for the SW transformation. Notably, such an expansion has removed the lowest-order coupling between the subsystems $\hat{V}$.

As an ansatz, we can let
\begin{equation}
    \hat{A} = \int d\mathbf{k} f(\mathbf{k})\hat{S}_{\mathbf{k}}\hat{R}^{\dagger}_{\mathbf{k}} - \text{H.c.},
\end{equation}
where $f(\mathbf{k})$ is an unknown function which should be chosen to satisfy~\eqref{eq:comm_req}.
Using~\eqref{eq:S_comm}, it is straightforward to show that we should choose $f(\mathbf{k})$ to take the form
\begin{equation}\label{eq:f_def}
    f(\mathbf{k}) = \frac{g_0}{\mathcal{E}_{\rm R}(\mathbf{k}) - \mathcal{E}_{\rm S}(\mathbf{k})}.
\end{equation}
We can now see from~\eqref{eq:f_def} that the perturbative expansion will be valid for dispersive couplings where $f(\mathbf{k}) \sim \epsilon$ for some small parameter $\epsilon$; otherwise, if $\mathcal{E}_{\rm S} \sim \mathcal{E}_{\rm R}$ for any values of $\mathbf{k}$ (i.e., if the interaction is resonant), then $f(\mathbf{k})$ becomes large and the expansion~\eqref{eq:SW_simpl} will not be accurate as an asymptotic expansion.

Assuming $f(\mathbf{k})$ to remain small such that $f(\mathbf{k}) \sim \epsilon$ for all values of $\mathbf{k}$, the leading-order term in $\epsilon$ in the SW expansion~\eqref{eq:SW_simpl} is $\frac{1}{2}[\hat{A},\hat{H}_0] =\hat{H}_{\rm S}^{(1)}$, where 
\begin{equation}
\label{eq:sw-int}
 \hat{H}_{\rm S}^{(1)} =  g_0 \iint d\mathbf{k}d \mathbf{k'}f(\mathbf{k})\left[\hat{S}_\mathbf{k}\hat{R}^{\dagger}_{\mathbf{k}}, \hat{S}^{\dagger}_{\mathbf{k}'}\hat{R}_{\mathbf{k}'}\right] + \text{H.c.}
\end{equation}

While the effective interaction \eqref{eq:sw-int} already provides new insight to the system behavior, we can further proceed with the elimination of the reservoir degrees of freedom using a \emph{mean-field approximation}. The mean field approximation consists of the leading-order approximation to the quantum dynamics of two coupled quantum subsystems~\cite{Degenfeld-Schonburg2014Jun}, and consists of neglecting any entanglement between the two systems and simply assuming $\ket{\psi} \approx \ket{\psi}_{\rm S}\otimes \ket{\psi}_{\rm R}$ (or the corresponding density operator relation for mixed states) for all times. As the two subsystems only interact to order $\sim f(\mathbf{k})$, we can expect any corrections beyond the mean-field to be subleading terms. In later sections of this paper, we investigate the limits of this approximation more carefully.

Under this approximation, we can perform a partial trace over the reservoir operators, which results in a reduced system Hamiltonian $\hat{H}'_{\rm S} = \hat{H}_{0}^{\rm S} + \hat{H}_{\rm S}^{(1)'}$, where $\hat{H}_{\rm S}^{(1)'} = \text{Tr}_{\rm R}\left[\hat{H}_{\rm S}^{(1)}\right]$. Finally, we often make the approximation that the dynamics of the reservoir under the mean-field approximation can be effectively approximated to leading order by simply $\hat{H}_{\rm R}' \approx \hat{H}_0^{\rm R}$. That is, for the purpose of obtaining the reduced system dynamics, we can treat the reservoir as independent of the system, to leading order. This is in spirit with the Born-Markov ME, where, when calculating \emph{system}-level observables, the reservoir state can be kept as its initial state to second order in the interaction between the subsystems~\cite{breuer2002}. 

Under this approximation, we can (often analytically) solve for the dynamics of the reservoir and insert them into the effective Hamiltonian $\hat{H}_{\rm S}'$. As an example, let us consider the case where $\hat{R}_{\mathbf{k}}$ is a product of annihilation operators of the reservoir band for momenta spanning the full space from $k_1$ to $k_N$.  If the reservoir begins in vacuum and remains sufficiently near so under the above approximation, we have, by Wick's theorem
\begin{equation}
    \langle \hat{R}_{\mathbf{k}} \hat{R}^{\dagger}_{\mathbf{k}'}\rangle = \sum_{\text{permutations of} \ \mathbf{k}} \delta(\mathbf{k}-\mathbf{k'})
\end{equation}
The Hamiltonian in this specific case then becomes
\begin{equation}
    \hat{H}_{\rm S}^{(1)'} = -g_0 N \int d\mathbf{k} f(\mathbf{k})\hat{S}^{\dagger}_{\mathbf{k}}\hat{S}_{\mathbf{k}}.
\end{equation}
Consequently, we obtain an effective model that is composed solely of system degrees of freedom.

\subsection{Dissipative and dispersive interactions - Born-Markov ME}
For the application of the SW transformation, it was essential that the function \eqref{eq:f_def} remained small for all $\mathbf{k}$, which we referred to as a dispersive interaction. In the case where $\mathcal{E}_{\rm R} - \mathcal{E}_{\rm S}\sim0$ for some values of $\mathbf{k}$ we have a phase-matching condition, which leads to resonant interaction. Certainly in this case, the dispersive approach of using a SW transformation breaks down for these regions of interaction space. Interestingly, this condition has a geometric interpretation as the real solutions to the equation
\begin{equation}\label{eq:curve}
   \mathcal{E}_{\rm R}(\mathbf{k}) - \mathcal{E}_{\rm S}(\mathbf{k}) =0,
\end{equation}
which forms a $(N-1)$-dimensional surface $\mathcal{S}$ in $\mathbb{R}^{N}$; for example, for the three-wave mixing (TWM) $\chi^{(2)}$ interaction studied in later sections of this work, this corresponds to 1-dimensional curve. In the common case where each band's dispersion relation is expanded up to second order in wavevector,~\eqref{eq:curve} corresponds to quadric hypersurfaces (conic sections for the TWM case).

An approach more suitable for dealing with resonant perturbative interactions is the Born-Markov ME approach~\cite{breuer2002}. This approach is based on a separation of timescales between the effective memory kernel of the system-reservoir interaction, and the coupling between the system and its environment, which allows for a Markovian (memoryless) description of the interaction. After performing a partial trace over the reservoir bands, we are left with an equation of motion for the reduced system density matrix, which is non-unitary in the presence of resonant system-reservoir interactions.

Under this approach, we have the ME for the system:
\begin{equation}\label{eq:me_1}
    \frac{d}{dt}\hat{\rho}_{\rm S} = -i[\hat{H}_0^{\rm S},\hat{\rho}_{\rm S}] + \mathbb{L}\hat{\rho}_{\rm S},
\end{equation}
where
\begin{equation}
    \mathbb{L}\hat{\rho}_{\rm S} = \int_0^{\infty} d\tau \text{Tr}_{\rm R}\left[\hat{\tilde{V}}(-\tau)\hat{\rho}_{\rm S} \hat{\rho}_{\rm R}, \hat{V}\right] + \text{H.c.},
\end{equation}
and $\hat{\tilde{V}}(t) = e^{i\hat{H}_0t}\hat{V}e^{-i\hat{H}_0t}$.
A derivation of this equation and an explanation of the approximations involved is given in Appendix.~\ref{app:master_equation}.

For the sake of illustration, we again can examine the example case where $\hat{R}_{\mathbf{k}}$ is a product of annihilation operators spanning the interaction space. Equation~\eqref{eq:me_1} then simplifies to 
    \begin{align}
    \frac{d}{dt}\hat{\rho}_{\rm S} &= -i[\hat{H}_0^{\rm S} + \hat{H}_{\rm S}^{(1)'},\hat{\rho}_{\rm S}] \nonumber \\ & + \int_{\mathcal{S}} d\mathbf{\Omega}_{\mathbf{k}} \left(\hat{L}_{\mathbf{k}}\hat{\rho}_{\rm S}\hat{L}^{\dagger}_{\mathbf{k}} - \frac{1}{2}\{\hat{L}^{\dagger}_{\mathbf{k}}\hat{L}_{\mathbf{k}},\hat{\rho}_{\rm S}\}\right)
\end{align}
where
\begin{equation}
    \hat{L}_{\mathbf{k}} = \frac{g_0^2 N\pi}{2} \frac{\hat{S}_{\mathbf{k}}}{|{\bm \nabla}\left[\mathcal{E}_{\rm R}(\mathbf{k}) - \mathcal{E}_{\rm S}(\mathbf{k})\right]|}
\end{equation}
is a Lindblad operator which depends on the continuous indices $\mathbf{k}$. The integration over the Lindblad operators is over the ``dissipation surface'' $\mathcal{S}$ defined by~\eqref{eq:curve}, if it exists. Otherwise, there is no dissipative component.

In fact, we only expect continuous Lindblad operators of this type in the case where the reservoir operators span the entire $N$-dimensional $\mathbf{k}$ space; for example, if one treats (part of) the FH as the reservoir, in TWM, which is done in Sec.~\ref{sec:dissipative_coupling}. In other situations, we should expect a non-Lindblad ME, which is due to the general inapplicability of the ``secular approximation" (sometimes called post-trace rotating-wave approximation) which is typically needed to put MEs in Lindblad form after the Born-Markov approximation in the case of continuous system degrees of freedom. It should be noted, however, that recent techniques should allow for casting the ME in Lindblad form if so desired~\cite{Nathan2020Sep}.

\begin{figure*}
    \includegraphics[width=0.9\textwidth]{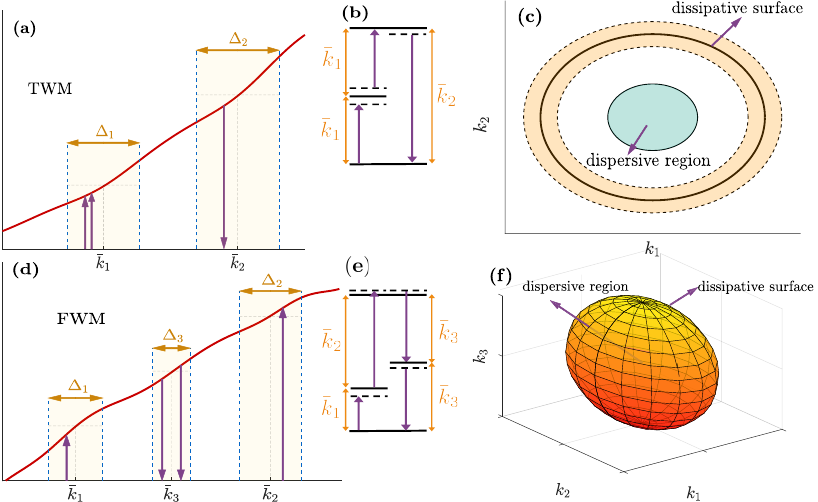}
    \caption{Schematic of interaction Hamiltonian and resonance condition geometry for examples of (a-c) three-wave mixing (TWM) and (d-f) four-wave mixing (FWM) for one set of band partitions. In (a,d), the overall energy band function $E(k)$ is shown schematically with band partitions with width $\Delta_i$ around center wavevector $\bar{k}_i$. (b,e) gives the corresponding energy-level diagram (neglecting dispersion for visual simplicity). Right panels show the geometry of the phase-matching resonance condition for (c) TWM with elliptical phase-matching solutions, where the orange annulus represents a region with (near)-resonant TWM where $\mathcal{E}_{\rm R}(k_1,k_2) \approx \mathcal{E}_{\rm R}(k_1,k_2)$, which can lead to exponential photon decay, and the blue inner region far from the resonance surface can (depending on the dispersion parameters) corresponds to a dispersive coupling regime where $f(k_1,k_2) \sim \epsilon$. (f) shows the same schematic for FWM with an ellipsoid phase-matching resonance condition, where the interior of the ellipsoid can in some instances correspond to a dispersive coupling regime in $\mathbf{k}$-space.
    }
    \label{fig:schematic1}
\end{figure*}

In Fig.~\ref{fig:schematic1}, we give a schematic of our formalism for the physical cases of three-wave mixing (TWM) and four-wave mixing (FWM), as well as an illustration of possible corresponding dissipation surfaces.

We see that the ME approach recovers, to leading order, the SW result under appropriate assumptions on the form of the reservoir operators and initial states. In general, the two approaches have their own pros and cons, however; the SW transformation allows for a more easy systematic expansion to higher orders of perturbation theory, and also gives more ability to model and track the dynamics of the bands we have called ``reservoirs" simultaneously with the system. The ME approach, on the other hand, is conceptually simpler in that it does not involve unitary transformation to a different frame (and thus one does not need to account for the change in initial conditions and observables this induces), and it is also able to handle dissipative interactions, whereas the SW can not. Both approaches also have their own merits in terms of gaining insight into the underlying physical processes. In fact, a combination of the two approaches can be more powerful than either of them individually, as we show in our analysis of the $\chi^{(2)}$ nonlinearity in Sec.~\ref{sec:sw}.

We  remark that there exist other techniques for eliminating degrees of freedom which could be incorporated into this multimode framework based on various types of adiabatic elimination (e.g., the effective operator formalism~\cite{Reiter2012Mar}, the ``dissipation-picture'' approach~\cite{Cirac1992Oct}, or quantum stochastic differential equation limit theorems~\cite{Mabuchi2009Oct}); for some of these formalisms, however, constructing higher-order contributions can be challenging.

\section{Quantum model for broadband $\chi^{(2)}$-nonlinear waveguide}\label{sec:model}

From this point, we apply the general formalism introduced in the previous section Sec.~\ref{sec:general} to a paradigmatic example of broadband cascaded $\chi^{(2)}$ nonlinear interactions. Our starting point is the Hamiltonian for a $\chi^{(2)}$-nonlinear waveguide
\begin{subequations}\label{eq:G_Hamiltonian}
\begin{align}
\hat{H}=\hat{H}_\mathrm{L}+\hat{H}_\mathrm{NL}
\end{align}
with
\begin{align}
\hat{H}_\mathrm{L}=\theta\int^\infty_{-\infty}dp\left(\frac{p^2}{2}\hat{\phi}_p^\dagger\hat{\phi}_p+\left(-\xi+\gamma p+\frac{\beta p^2}{2}\right)\hat{\psi}_p^\dagger\hat{\psi}_p\right)
\end{align}
 and
 \begin{align}
\hat{H}_\mathrm{NL}=\frac{1}{2}\int^\infty_{-\infty}dpdq\left(\hat{\psi}_p^\dagger\hat{\phi}_{\frac{p}{2}+q}\hat{\phi}_{\frac{p}{2}-q}+\hat{\psi}_p\hat{\phi}^\dagger_{\frac{p}{2}+q}\hat{\phi}^\dagger_{\frac{p}{2}-q}\right).
\end{align}
\end{subequations}
Note that we have scaled the time and space appropriately to normalize the Hamiltonian to a dimensionless form (see Appendix~\ref{app:nondimensionalization} for the nondimensionalization procedure). Here, $\hat{\phi}_p$ and $\hat{\psi}_p$ are annihilation operators for FH and SH photons with normalized wavenumber $p$. The dispersion relation $\omega(k)$ is expanded to second order around FH center-wavevector $k_0$ and SH center-wavevector $2k_0$.
$\theta$ is the sign of signal group-velocity dispersion, $\xi$ is a normalized phase-mismatch, $\gamma$ is a normalized group-velocity mismatch, and $\beta$ is a normalized group-velocity dispersion.

In the language of Sec.~\ref{sec:general}, the interaction space is given by the momentum vector $\mathbf{K}=(p,q)$, with $\hat{S}_{\mathbf{K}} = \hat{\phi}_{\frac{p}{2}+q}\hat{\phi}_{\frac{p}{2}-q}$, and $\hat{R}_{\mathbf{K}} = \hat{\psi}_{p}$, and $g_0 =1/2$. 

The nonlinear term $\hat{H}_\mathrm{NL}$ denotes an interaction via which a single SH photon with momentum $p$ is annihilated to generate a pair of FH photons with momentum $\frac{p}{2}\pm q$, and vice versa, where $p$ and $q$ are arbitrary real numbers. The linear energy (i.e., from $\hat{H}_\mathrm{L}$) of the SH photon and the pair of FH photons are given as
\begin{align}
\begin{split}
    &\mathcal{E}_\text{SH}(p)=\theta\left(-\xi+\gamma p+\frac{\beta p^2}{2}\right)\\
    &\mathcal{E}_\text{FW}(p,q)=\theta\left(\frac{p^2}{4}+q^2\right),
\end{split}
\end{align}
respectively. Here, $\mathcal{E}_{\rm FW}(p,q) = \mathcal{E}_{\rm S}(\mathbf{K})$, and $\mathcal{E}_{\rm SH}(p) = \mathcal{E}_{\rm R}(\mathbf{K})$.

\begin{figure}
    \includegraphics[width=0.48\textwidth]{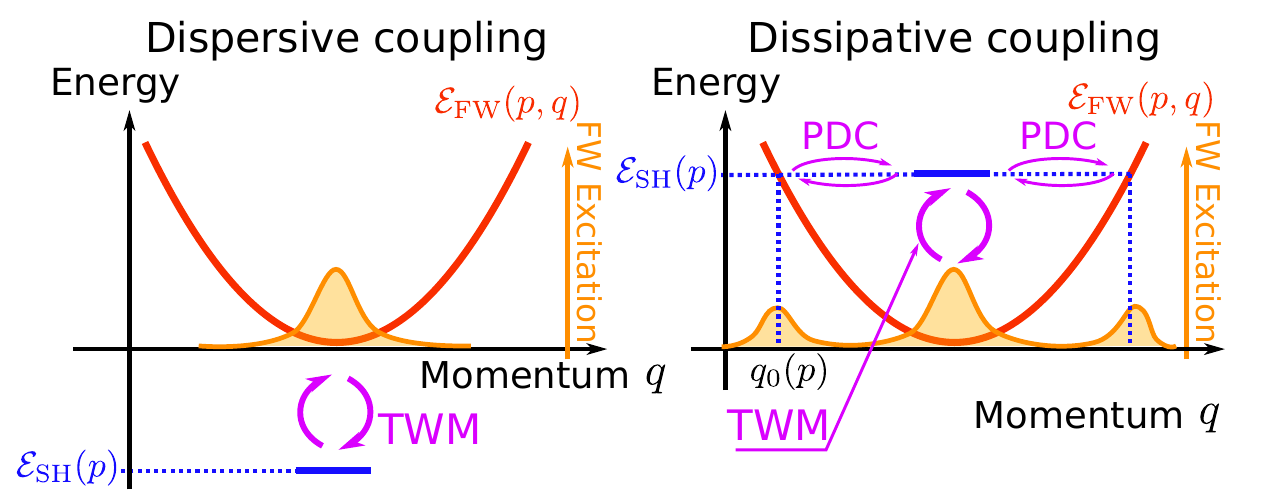}
    \caption{Energy of an SH photon $\mathcal{E}_\text{SH}(p)$ with momentum $p$ is shown together with the energies of a pair of FH photons $\mathcal{E}_\text{FW}(p,q)$ with momenta $\frac{p}{2}\pm q$. TWM mediates interactions between FH and SH (purple arrows). Orange envelopes represent typical FH excitations as a result of system dynamics. Left: In the dispersive coupling regime, the TWM process is off-resonant for any combination of $p$ and $q$. Right: In the dissipative coupling regime, there exists $q_0(p)$ that makes $\mathcal{E}_\text{SH}(p)\approx \mathcal{E}_\text{FW}(p,q_0(p))$. As a result, a parametric downconversion (PDC) process from the SH then populates FH modes around $\frac{p}{2}\pm q_0(p)$.}
    \label{fig:energylevels}
\end{figure}

We refer to the nonlinear interaction via $\hat{H}$ as ``dispersive'' when
\begin{align}
    \label{eq:dispersive_condition}
    |\mathcal{E}_\text{FW}-\mathcal{E}_\text{SH}|\sim |\xi| \gg 1
\end{align}
holds true for any combination of $p$ and $q$. In this regime, we can assume that the nonlinear interaction is always highly phase mismatched as schematically shown in Fig.~\ref{fig:energylevels}. As we show in later sections, this condition ensures the validity of perturbative treatments based on an expansion parameter $\xi^{-1}$ for the entire Hilbert space, allowing us to derive concise analytic results. In terms of the system parameters, \eqref{eq:dispersive_condition} can be reformulated as 
\begin{align}
    \label{eq:dispersive_inequalities}
    &\beta<\frac{1}{2}, &\xi \gg1.
\end{align}
Notice that this condition enforces the sign of the phase-mismatch to positive $\xi>0$, which fixes the relative sign between group velocity dispersion and nonlinear interactions. We address the dispersive coupling regime in Sec.~\ref{sec:dispersive_coupling} in detail.

When the system parameters do not fulfill \eqref{eq:dispersive_inequalities} the nonlinear interaction is close to resonance for a certain combination of $p$ and $q$ as depicted in Fig.~\ref{fig:energylevels}. In particular, the case with negative phase mismatch $\xi \ll -1$ is of most interest, since it allows an access to cubic nonlinearity with opposite sign, and we refer to this regime as the ``dissipative'' coupling regime. Special care has to be taken in the vicinity of the resonance, where perturbative treatment of the nonlinearity breaks down, and we provide a detailed analysis in Sec.~\ref{sec:dissipative_coupling}. It is important to note that even in the dissipative coupling regime, the effective cubic cascaded nonlinearity arises from a dispersive interaction between low-momentum FH and SH modes---the ``dispersive'' coupling regime thus should be understood as referring to interactions which are dispersive across the \emph{entire} interaction band.

\section{Classical Derivation of effective cubic nonlinearity}\label{sec:classical}
Upon introducing annihilation operators in the normalized spatial coordinate $y=z/z_c$ as
\begin{subequations}
    \begin{align}
        \label{eq:spatialoperators_lab}
        \hat{\Phi}_y&=\int^\infty_{-\infty}dpe^{-2\pi ipy}\hat{\phi}_p, &\hat{\Psi}_y=\int^\infty_{-\infty}dpe^{-2\pi ipy}\hat{\psi}_p\\
        \hat{\phi}_p&=\int^\infty_{-\infty}dye^{2\pi ipy}\hat{\Phi}_y, &\hat{\psi}_p=\int^\infty_{-\infty}dye^{2\pi ipy}\hat{\Psi}_y,
    \end{align}
    \end{subequations}
the Hamiltonian \eqref{eq:G_Hamiltonian} is equivalently given as
\begin{align}
    \label{eq:spatialhamiltonian_lab}
    \hat{H}=&\int_{-\infty}^{\infty}dy\left[-\frac{\theta}{2}\hat{\Phi}_y^\dagger\slashed{\partial}_y^2\hat{\Phi}_y-\theta\xi\hat{\Psi}_y^\dagger\hat{\Psi}_y+ i\theta\gamma\hat{\Psi}_y^\dagger\slashed{\partial}_y\hat{\Psi}_y\right.\nonumber\\
    &\quad{}\left.-\frac{\theta\beta}{2}\hat{\Psi}_y^\dagger\slashed{\partial}_y^2\hat{\Psi}_y+\frac{1}{2}\left(\hat{\Psi}_y^\dagger\hat{\Phi}_y^2+\hat{\Phi}_y^{
    \dagger 2}\hat{\Psi}_y\right)\right],
\end{align}
where $\slashed{\partial}_y=\partial_y/2\pi$.

Staring from \eqref{eq:spatialhamiltonian_lab}, we sketch a classical derivation~\cite{Menyuk1994} of an effective cubic Hamiltonian in the large phase-mismatch regime $|\xi|\gg 1$, which provides a useful reference for the later quantum mechanical treatment. Assuming the classical mean-field limit, we substitute operator-valued fields with c-number fields as $\hat{\Phi}_y\mapsto\Phi_y$ and $\hat{\Psi}_y\mapsto\Psi_y$ in the Heisenberg equations of motion of \eqref{eq:spatialhamiltonian_lab}, which leads to expressions equivalent to classical coupled-wave equations:
\begin{subequations}
\begin{align}
    \label{eq:phiclassical}
     i\frac{d\Phi_y}{dt}&=-\frac{\theta}{8\pi^2}\partial_y^2\Phi_y+\Psi_y\Phi_y^*\\
    \label{eq:psiclassical}
     i\frac{d\Psi_y}{dt}&=-\theta\xi\Psi_y+\frac{ i\theta\gamma}{2\pi}\partial_y\Psi_y-\frac{\theta\beta}{8\pi^2}\partial_y^2\Psi_y+\frac{1}{2}\Phi_y^2.
\end{align}
\end{subequations}
Assuming $\xi$ dominates over other energy scales, from \eqref{eq:psiclassical} one can see that $\Psi_y$ will rapidly oscillate at frequency $\sim \xi$. Neglecting (i.e., averaging over) these fast oscillations,  we can approximate $ \frac{d\Psi_y}{dt} \approx 0$ to obtain 
\begin{align}
    \label{eq:phiadiabatic}
    \Psi_y\approx \frac{\theta}{2\xi}\Phi_y^2
\end{align}
from \eqref{eq:psiclassical} to the lowest order in $\xi^{-1}$. We adiabatically eliminate the SH modes by inserting \eqref{eq:phiadiabatic} into \eqref{eq:phiclassical}, which leads us to a nonlinear Schr\"odinger equation (NLSE) for the FH fields
\begin{align}
    \label{eq:classicalnlse}
     i\frac{d\Phi_y}{dt}&\approx-\frac{\theta}{8\pi^2}\partial_y^2\Phi_y+\frac{\theta}{2\xi}\Phi_y^*\Phi_y^2.
\end{align}
Moving back to a quantum picture, one can deduce the form of an underlying cubic Hamiltonian as
\begin{small}
\begin{align}
    \label{eq:classical_cubic}
    &\hat{H}_{\text{cubic}}=\theta\int_{-\infty}^{\infty}dy\left(-\frac{1}{2}\hat{\Phi}_y^\dagger\slashed{\partial}_y^2\hat{\Phi}_y+\frac{1}{4\xi}\hat{\Phi}_y^{\dagger 2}\hat{\Phi}_y^2\right)\\
    =&\theta\int^\infty_{-\infty} \! dp\left(\frac{p^2}{2}\hat{\phi}_p^\dagger\hat{\phi}_p+\frac{1}{4\xi}\iint\! dqdq'\hat{\phi}_{\frac{p}{2}+q}^\dagger\hat{\phi}_{\frac{p}{2}-q}^\dagger\hat{\phi}_{\frac{p}{2}+q'}\hat{\phi}_{\frac{p}{2}-q'}\right)\nonumber,
\end{align}\end{small}which describes a dispersionless SPM. For $\theta=1$, $\xi>0$ and $\xi<0$ lead to repulsive and attractive interactions, respectively.
\section{Schrieffer-Wolff Transformation Approach}\label{sec:sw}
In this section, we present a derivation of the cascaded $\chi^{(2)}$ quantum nonlinearity for a multimode system by using the Schrieffer–Wolff (SW) transformation~\cite{Schrieffer1966,Bravyi2011} to analyze
\eqref{eq:G_Hamiltonian}, where the inverse of the phase-mismatch, $\xi^{-1}$, is used as an expansion parameter. 
\subsection{Dispersive coupling regime}
\label{sec:dispersive_coupling}
In this subsection we focus on the dispersive coupling regime. Here, in the interaction frame given by the SW transformation, the dressed FH modes experience self-phase modulation, as expected from the classical results, in addition to cross-phase modulation by the dressed SH modes. Analysis reveals dressed SH quanta as optical mesons~\cite{Drummond1997}, as shown in Appendix~\ref{app:mesons}. Through full quantum simulation, we confirm the convergence of the  dynamics of FH in the lab frame under \eqref{eq:G_Hamiltonian} to the heuristic Hamiltonian \eqref{eq:classical_cubic} as $\xi\rightarrow\infty$, and discuss the order in perturbation theory to which the cascade Hamiltonian is valid.
\subsubsection{Hamiltonian perturbation}
Following the prescription of SW transformations, let us consider a unitary transformation $\hat{U}=e^{\hat{S}}$, where $\hat{S}$ is an anti-Hermitian operator. In the SW frame given by $\hat{U}$, the Hamiltonian becomes
\begin{align}
    \hat{H}'=\hat{U}\hat{H}\hat{U}^\dagger.
\end{align}
We then assume that we have found an $\hat{S}$ that fulfills 
\begin{align}
    \label{eq:swcondition}
    [\hat{H}_{\rm L},\hat{S}]=\hat{H}_{\rm NL}.
\end{align}
As $\hat{H}_{\rm L}\sim\xi$ and $\hat{H}_{\rm NL}\sim 1$, we expect $\hat{S}\sim\xi^{-1}$, and as a result, we obtain
\begin{align}
    \label{eq:bchexpansion}
    \hat{H}'=\hat{H}_{\rm L}+\frac{1}{2}[\hat{S},\hat{H}_{\rm NL}]+\mathcal{O}(\xi^{-\frac{3}{2}}),
\end{align}
where $\hat{H}_{\rm NL}$ has been canceled by $[\hat{S},\hat{H}_{\rm L}]=-\hat{H}_{\rm NL}$. Although one might expect the next-to-leading order term to scale as $\xi^{-2}$, in fact, the multimode nature of the expansion means it in fact scales as $\xi^{-\frac{3}{2}}$, which we show in Appendix~\ref{app:higherorder}. Note that operators $\hat{\phi}_s$ and $\hat{\psi}_s$ appearing in expressions for $\hat{H}'$ annihilate excitations in the interaction frame, and these excitations correspond to ``dressed'' FH and SH modes in the lab frame. To find a solution of \eqref{eq:swcondition}, we posit the form of $\hat{S}$ as
\begin{align}
    \label{eq:swoperator}
    \hat{S}=\iint^\infty_{-\infty}dpdqf(p,q)\left(\hat{\psi}_p^\dagger\hat{\phi}_{\frac{p}{2}+q}\hat{\phi}_{\frac{p}{2}-q}-\hat{\psi}_p\hat{\phi}^\dagger_{\frac{p}{2}+q}\hat{\phi}^\dagger_{\frac{p}{2}-q}\right)
\end{align}
and solve the equation \eqref{eq:swcondition} for $f(p,q)$. Consistent with Eq.~\eqref{eq:f_def}, we obtain
\begin{align}
    \label{eq:fsq}
    &f(p,q)=\frac{1}{2(\mathcal{E}_\text{SH}-\mathcal{E}_\text{FW})}\\
    &=-\frac{\theta}{2}\left(\frac{1}{4}(1-2\beta)p^2-\gamma p+\xi+q^2\right)^{-1}\sim\xi^{-1}\nonumber
\end{align}
as a solution, where the condition \eqref{eq:dispersive_condition} ensures that $f(p,q)$ does not diverge. The leading-order scaling $f(p,q) \sim \xi^{-1}$ applies when the dominant system excitations remain bounded such that  $p^2$ and $q^2$ are smaller than $\sim\xi$.

Based on \eqref{eq:swoperator} and \eqref{eq:fsq}, we can evaluate the lowest order perturbation term in \eqref{eq:bchexpansion} as
\begin{align}
    \frac{1}{2}[\hat{S},\hat{H}_{\rm NL}]=\hat{W}_\text{lin}+\hat{W}_\text{XPM}+\hat{W}_\text{SPM}
\end{align}
where
\begin{align}
    \hat{W}_\text{lin}=-\frac{\pi\theta}{2}\int^\infty_{-\infty}dp\frac{\Theta(\omega(p))}{\sqrt{\omega(p)}}\hat{\psi}_p^\dagger\hat{\psi}_p
\end{align}
is a linear energy shift on the dressed SH modes with 
\begin{align}
    \label{eq:omega}
   \omega(p)=\frac{p^2}{4}-\theta\mathcal{E}_\text{SH}(p),
\end{align}
and $\Theta(\omega)$ is the Heaviside step function. In the dispersive regime, $\omega(p) \geq 0$ whenever $ \gamma^2/(1-2\beta) \leq \xi$ --- thus, except for $\beta \approx 1/2$, $\omega$ is always positive when $|\gamma| \ll \xi$, which we assume to be the case for the rest of the paper for simplicity. 

The nonlinear terms are:
\begin{align}
    \hat{W}_\text{SPM}=&-\frac{1}{4}\iiint^\infty_{-\infty}\!\!\! dpdqdrf(p,q)\hat{\phi}_{\frac{p}{2}+r}^\dagger\hat{\phi}_{\frac{p}{2}-r}^\dagger\hat{\phi}_{\frac{p}{2}+q}\hat{\phi}_{\frac{p}{2}-q}\nonumber\\
    &\quad{}+\mathrm{H.c.},
\end{align}
which represents SPM, and
\begin{align}
    \hat{W}_\text{XPM}=&\iiint^\infty_{-\infty}dpdqdrf(p,q)\hat{\psi}_p^\dagger\hat{\psi}_r\hat{\phi}^\dagger_{r-\frac{p}{2}-q}\hat{\phi}_{\frac{p}{2}-q}\nonumber\\
    &+\mathrm{H.c.},
\end{align}
corresponds to cross-phase modulations (XPM). The nonvanishing dependence of $f(p,q)$ on momenta $p$ and $q$ represents the finite spatial correlation length ($\sim \sqrt{\xi})$ of these nonlinear interactions. Combining these results, we obtain
\begin{align}
    \label{eq:lowestperturbation}
    \hat{H}'=\hat{H}_{\rm L}+\hat{W}_\text{lin}+\hat{W}_\text{XPM}+\hat{W}_\text{SPM}+\mathcal{O}(\xi^{-\frac{3}{2}})
\end{align}

For initial states sufficiently localized in momentum space, one may assume that the FH and SH bandwidth is always limited throughout the system evolution such that 
\begin{align}
    \label{eq:narrowband}
    f(p,q)\approx -\theta/2\xi + \mathcal{O}(\xi^{-2})
\end{align}
holds true for relevant values of $p$ and $q$. Heuristically, this condition is met when the magnitude of the normalized momentum of each photon remains smaller than $\sim\sqrt{\xi}$. When \eqref{eq:narrowband} is a good approximation, the SPM induced by $\hat{H}'$ becomes equivalent to that of $\hat{H}_\text{cubic}$, and we obtain an effective Hamiltonian in the spatial domain as
\begin{align}
    \label{eq:lieb}
    \begin{split}
    \hat{H}'\approx&\theta \int^\infty_{-\infty}dy\left(-\frac{1}{2}\hat{\Phi}_y^\dagger\slashed{\partial}_y^2\hat{\Phi}_y+\hat{\Psi}_y^\dagger M( i\slashed{\partial}_y)\hat{\Psi}_y\right.\\
&+\left.\xi^{-1}\hat{\Psi}_y^\dagger\hat{\Psi}_y\hat{\Phi}_y^\dagger\hat{\Phi}_y+\frac{1}{4}\xi^{-1}\hat{\Phi}_y^{\dagger 2}\hat{\Phi}_y^2\right),
    \end{split}
\end{align}
where $M(p) = \mathcal{E}_{\rm SH}(p) -\frac{\pi \theta}{2\sqrt{\omega(p)}}$.
When the input state contains a wider band of momentum states, the full expression of \eqref{eq:lowestperturbation} has to be employed, which does not necessarily have a convenient form in the spatial domain.

It is worth mentioning that, up to the order of perturbation we consider here, both the dressed FH population
and the dressed SH population
are conserved quantities. In other words, the unitary transformation $\hat{U}=e^{\hat{S}}$ partitions the entire Hilbert space into two sectors, i.e., dressed FH and dressed SH, within which particle numbers are approximately conserved.

Finally, we note that if the initial state of the system has no photons in the SH modes, then it is possible to show that~\eqref{eq:lowestperturbation} in fact remains accurate up to $\mathcal{O}(\xi^{-\frac{5}{2}})$, which we demonstrate in Appendix~\ref{app:higherorder}.

As, to order $\xi^{-1}$ in the SW expansion, the SW frame Hamiltonian is diagonal in the SH and FH operators, we expect these dressed operators to correspond to bound quasiparticle ``optical mesons'' excitations in the lab frame, which we discuss in Appendix~\ref{app:mesons}.

\subsubsection{Numerical simulations}
\label{sec:dispersive_simulation}
For large $\xi\gg1$, the SW transformation is close to the identity, i.e., $\hat{U}\approx \hat{I}$, and thus, dressed FH and SH excitations are also approximately equivalent to FH and SH excitations  in the lab frame. As a result, dynamics of FH modes under $\hat{H}$ in the absence of initial SH excitations are close to those under a cubic Hamiltonian $\hat{H}_\text{cubic}$, which we confirm numerically. Specifically, in Sec.~\ref{sec:mf} we show that observables calculated with the two different approaches should only differ at order $\mathcal{O}(\xi^{-2})$, which can be further improved by using the semi-analytic mean-field theory, or an adiabatic chirping of the phase-mismatch to ensure vacuum dressed SH modes. 

To perform numerical simulations, we discretize the reciprocal space Hamiltonian by introducing
\begin{align}
    &\hat{\phi}_p=\sqrt{L}\hat{a}_p, &\hat{\psi}_p=\sqrt{L}\hat{b}_p,
\end{align}
and letting $p \rightarrow  L/p$, where $p$ is now an integer and $L$ is the finite system size. The Hamiltonian then becomes
\begin{align}\label{eq:disc}
    \hat{H} &= \theta\sum_{p}\left[ \frac{(p/L)^2}{2}\hat{a}^{\dagger}_p \hat{a}_p + \left(-\xi + \gamma \frac{p}{L} + \frac{\beta (p/L)^2}{2}\right)\hat{b}^{\dagger}_p \hat{b}_p\right] \nonumber \\ 
    & + \frac{1}{2\sqrt{L}}\sum_{pq}\left(\hat{b}^{\dagger}_p \hat{a}_{q}\hat{a}_{p-q} + \hat{b}_p \hat{a}^{\dagger}_q \hat{a}^{\dagger}_{p-q} \right).
\end{align}

We also introduce a corresponding discretized cubic Hamiltonian
\begin{align}
    \label{eq:gcubic_discrete}
    \begin{split}
    \hat{H}_\text{cubic}&=\sum_p\frac{\theta p^2}{2L^2}\hat{a}_p^\dagger\hat{a}_p\\
    &+\frac{\theta}{4\xi L}\sum_{\substack{p_1+p_2=p_3+p_4}}\hat{a}_{p_1}^\dagger\hat{a}_{p_2}^\dagger\hat{a}_{p_3}\hat{a}_{p_4}.
    \end{split}
\end{align}

As an initial state, we consider the $p=0$ coherent state
\begin{align}
    \label{eq:coherent}
    \ket{\alpha}=e^{-\frac{|\alpha|^2}{2}}\sum_{n=1}\frac{\alpha^n}{\sqrt{n!}}\ket{n},
\end{align}
where $\ket{n}=\frac{1}{\sqrt{n!}}\hat{a}_0^{\dagger n}\ket{0}$. Note that $|\alpha|^2$ corresponds to the mean photon number in the quantization window with size $L$, and photon flux density becomes $\rho(y)=|\alpha|^2/L$. When $\alpha$ is chosen small enough, one can truncate the sum in \eqref{eq:coherent} at a finite $n$. For numerical simulations in this section, we choose $\alpha=0.1$ and truncate the sum at $n=2$. For larger photon number regimes, simulations could be done using (e.g.,) matrix product states~\cite{Yanagimoto2021Oct}.

\begin{figure}
    \centering
    \includegraphics[width=0.48\textwidth]{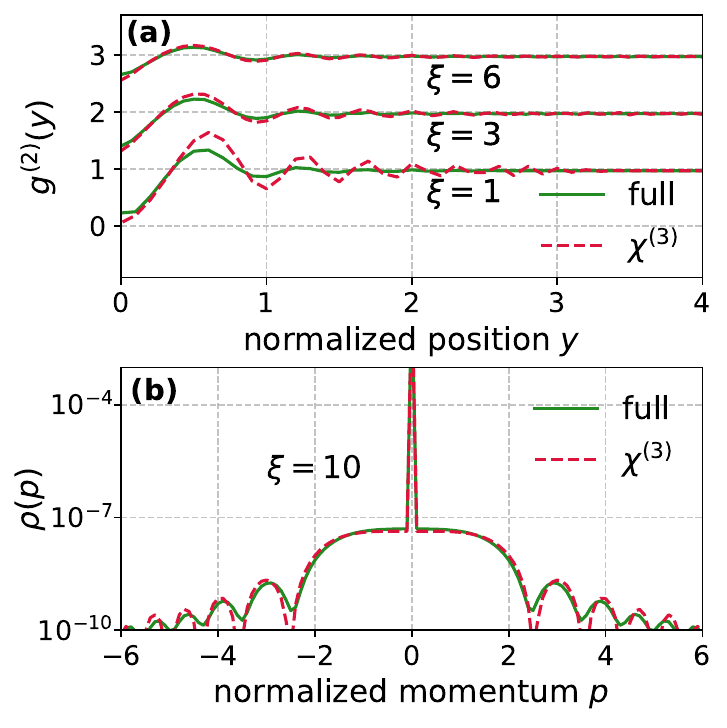}
    \caption{ (a) Two-photon spatial correlation function $g^{(2)}(y)$ and (b) momentum distribution $\rho(s)$ of
FH photons prepared initially in a coherent state of $\alpha = 0.1$ for momentum $p = 0$ in the dispersive coupling regime. The green solid lines show the full TWM Hamiltonian (discretized in~\eqref{eq:disc}) result, and the dashed red lines show the approximate SW result using the cubic Hamiltonian in Eq.~\eqref{eq:gcubic_discrete}. Here we let $\gamma = 0$ and $\beta = -1$, and show data at time $t=2$.} 
    \label{fig:repulsive2}
\end{figure}

In Fig.~\ref{fig:repulsive2}, we show the two-photon  correlation function
\begin{align}
    g^{(2)}(y)=\frac{\langle \hat{\Phi}^\dagger_y\hat{\Phi}^\dagger_0\hat{\Phi}_0\hat{\Phi}_y\rangle}{\rho(y=0)^2},
\end{align}
and momentum distribution
\begin{align}
    \rho(p)=\langle \hat{\phi}^\dagger_p\hat{\phi}_p\rangle
\end{align}
of the FH state after a finite time $t$ under both $\hat{H}$ and $\hat{H}_\text{cubic}$. The strong anti-bunching observed at $y=0$ followed by oscillations in $g^{(2)}(y)$ is a characteristic feature of fermionization of bosons in a quenching dynamics under repulsive interactions~\cite{Cantu2020,Muth2010,Muth2010_b}. For the simulation, we choose simulation times on the order of unity, which, when scaled by $1/g_c$ corresponds to an effective interaction length on an order of few meters for current state of the art nanophotonic waveguides~\cite{Yanagimoto2020_fano}. Agreement between the two dynamics confirms the validity of $\hat{H}_\text{cubic}$ as an approximate expression for the quantum propagation of FH even in this highly nonlinear regime. In fact, as we show in Sec.~\ref{sec:dissipative_coupling}, higher-order corrections are required in the case of the dissipative coupling regime to obtain agreement in the limit $|\xi| \rightarrow \infty$.
\subsubsection{Prospects for engineering spatial potential}
While we have mainly focused on the cubic nonlinearity that FH photons experience with vacuum input to the SH modes, varying the initial SH excitations could allow for access to even richer quantum dynamics. More specifically, we can assume a coherent state initial SH excitation with spatial variation
\begin{align}
    \langle\hat{\Psi}_y\rangle(t=0)=v(y)
\end{align}
in the lab frame. When $\xi\gg1$ is fulfilled, $\langle\hat{\Psi}_y\rangle\approx v(y)$ also holds true in the interaction frame to leading order, provided $|v(y)| \gg \xi^{-1}$. 
 As a result, we can perform a formal substitution $\hat{\Psi}_y\mapsto v(y)$, which leads us to, at time $t=0$,
\begin{align}
    \label{eq:lieb_spatial}
    \begin{split}
    \hat{H}'\approx&\theta \int^\infty_{-\infty}dy\left(-\frac{1}{2}\hat{\Phi}_y^\dagger\slashed{\partial}_y^2\hat{\Phi}_y-V(y)\hat{\Phi}_y^\dagger\hat{\Phi}_y+\frac{1}{4}\xi^{-1}\hat{\Phi}_y^{\dagger 2}\hat{\Phi}_y^2\right)
    \end{split}
\end{align}
where
\begin{align}
    V(y)=-\xi^{-1}|v(y)|^2.
\end{align}
Physically, for $\theta=1$, this Hamiltonian describes massive bosons with a repulsive delta-interaction under an additional spatial potential $V(y)$.

Additionally, we assume that the SH dispersion is specifically engineered to flatten $M(p)$, which ensures that the waveform of dressed SH excitation does not change significantly as time passes. For example, if $\gamma =0$, we can write the potential Hamiltonian term under the mean-field approximation (discussed in Sec.~\ref{sec:mf}) as
\begin{equation}
    -\frac{\theta}{2}\int_{-\infty}^{\infty} dy  V'(y,t) \hat{\Phi}^{\dagger}_{y} \hat{\Phi}_{y}  + \text{H.c.},
\end{equation}
where
\begin{equation}
V'(y,t) = \int_{-\infty}^{\infty} dy' K(y,y';t')V(y'),
\end{equation}
and $K(y,y';t')$ is the Wick-rotated heat kernel for imaginary time $t' = i8\pi^2 t/\beta$:
\begin{equation}
K(t',y,y') = \frac{1}{\sqrt{4\pi t'}}e^{-\frac{(y-y')^2}{4t'}},
\end{equation}
which satisfies $K(t'=0,y,y') = \delta(y-y')$. Thus, a potential with spatial extent $L_{\rm c}$ will remain coherent on a timescale $T_{\rm c} \sim L_{\rm c}^2/|\beta|$. The time evolution of $V'(y,t)$ is equivalent to the evolution of the wavefunction of a free quantum particle with mass $\propto \beta$ dispersing under the Schr{\"o}dinger equation from intial state $V(y)$.

It is worth mentioning that realizing such spatial potential is typically quite challenging for photons, since the potential has to be co-propagating at the speed of light; in contrast, cross-phase modulation provides a straightforward realization for our case. Notably, the presence of a spatial potential allows access to a broader class of 1D many-body Hamiltonians known to exhibit exotic features, including the Bose glass, Mott insulator,~\cite{Yao2020}, and Wigner crystallization~\cite{Otterbach2013} to name a few.
\subsubsection{Validity of SW expansion for Hamiltonian dynamics}
As the SW expansion Hamiltonian $\hat{H}'$ was derived using a perturbative expansion by means of unitary transformation to the SW frame, it is worth examining to what order in perturbation theory the expansion can be considered reliable. 

First, consider an initial condition of excited \emph{dressed SW frame} FH photon states, and vacuum for the dressed SW frame SH. Following the argument in Appendix~\ref{app:higherorder}, we know that the excitations in the dressed SH induced by the higher-order interactions with the dressed FH scale as $\hat{\psi} \sim \xi^{-\frac{5}{2}}$ to leading order. Thus, examining~\eqref{eq:lowestperturbation}, the leading-order correction due to induced dressed SH excitations scales as $\xi^{-4}$ (from the $\hat{H}_{\rm L}$ term), which can of course be neglected as it is of a higher order than the terms we consider in the SW expansion. The lowest-order term in the SW expansion we have neglected which contributes when the dressed SH modes are in vacuum scales as $\xi^{-\frac{5}{2}}$, as we also show in Appendix~\ref{app:higherorder}.

We thus conclude that for an initial condition of vacuum \emph{dressed} SH modes, the Hamiltonian in~\eqref{eq:lowestperturbation} is valid up to $\mathcal{O}(\xi^{-\frac{5}{2}})$. This can be slightly improved to $\mathcal{O}(\xi^{-3})$ by using $\hat{W}'_{\rm SPM}$ (see~\eqref{eq:Wprime}) in place of $\hat{W}_{\rm SPM}$. The cubic Hamiltonian~\eqref{eq:lieb} is valid to $\mathcal{O}(\xi^{-2})$; however, as the cubic Hamiltonian essentially only differs from $\hat{W}_{\rm SPM}$ and $\hat{W}'_{\rm SPM}$ \emph{quantitatively}, it is sufficient to use ~\eqref{eq:lieb} as $\xi \rightarrow \infty$ to obtain the correct dynamics and capture all important features.

The (generally more physical) case of initial vacuum for the \emph{lab frame} SH is discussed in Sec.~\ref{sec:mf}.

\subsection{Dissipative coupling regime}
\label{sec:dissipative_coupling}
\begin{figure}[bth]
    \includegraphics[width=0.48\textwidth]{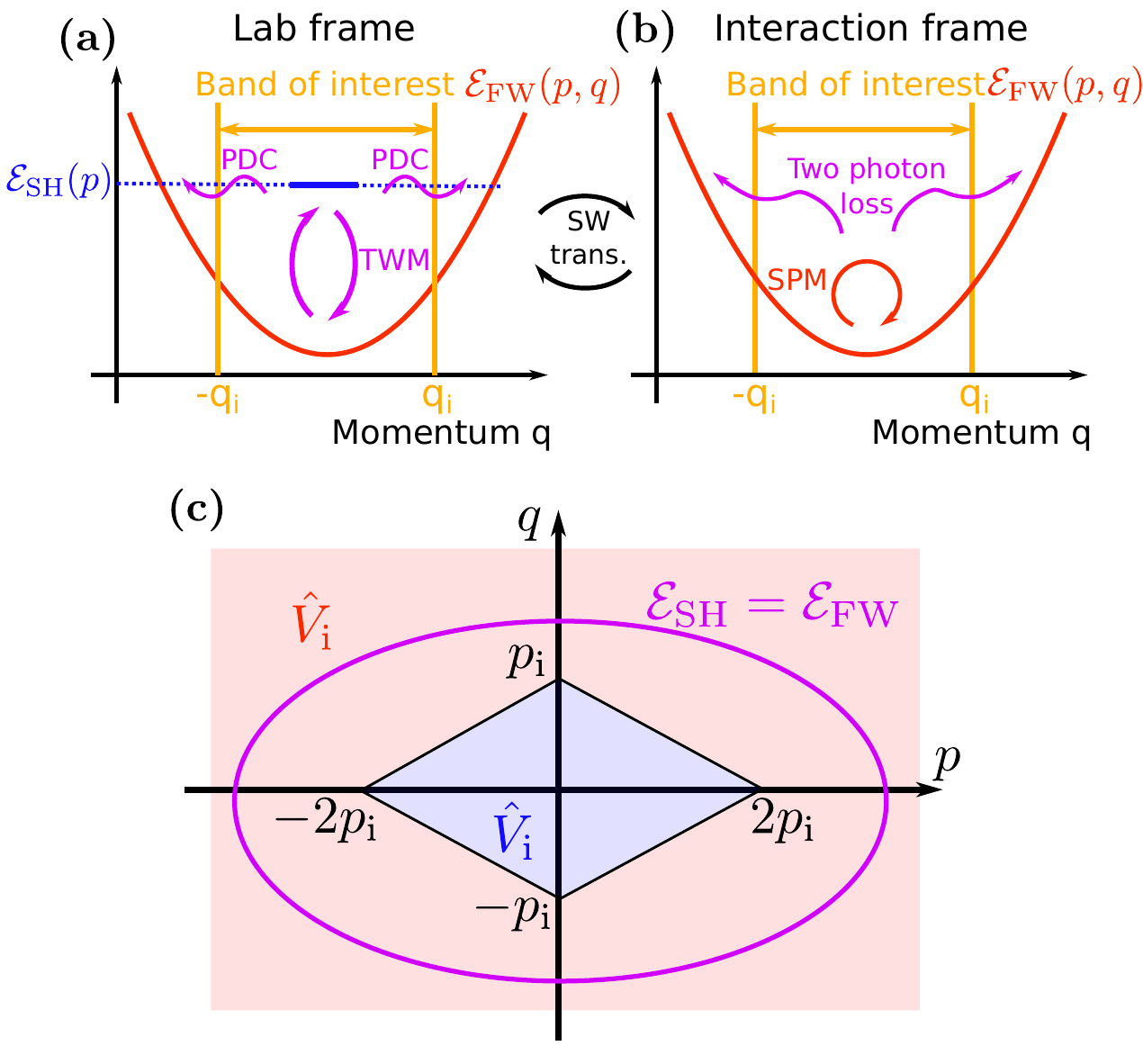}
    \caption{(a,b) Illustration of energy levels and interactions in the dissipative coupling regime. (a) In the lab frame, each SH photon is coupled resonantly with a continuum of FH states, leading to dissipative parametric downconversion (PDC). (b) In the SW frame, FH photons within the band of interest interact via SPM, and dissipative PDC is captured as two photon loss. (c) The TWM process is resonant when $\mathcal{E}_\text{SH}(p)=\mathcal{E}_\text{FW}(p,q)$ is fulfilled for the momentum of a SH photon $p$ and a pair of FH photons $\frac{p}{2}\pm q$, which forms an ellipse in the $(p,q)$-plane (purple oval, here $\gamma=0$). Region of integration for inside-the-band component $\hat{V}_\text{i}$ (blue shaded area) is taken to avoid the resonance, while $\hat{V}_\text{e}$ captures the rest.}
    \label{fig:dissipative_coupling}
\end{figure}

In the dissipative coupling regime, as illustrated in Fig.~\ref{fig:energylevels} and Fig.~\ref{fig:dissipative_coupling}(a), each SH photon with momentum $p$ is resonantly coupled to a continuum of FH modes with similar energy via the TWM interactions. Because of the resonant nature of the coupling, 
we cannot treat such a TWM process exclusively using the SW transformation approach. In fact, in the few-photon regime, SH photons exponentially decay into the FH continuum to form dispersive FH waves via a process analogous to atomic/molecular autoionization~\cite{Yanagimoto2020_fano}, as well as supercontinuum generation~\cite{Langford2011Oct}, showing that SH photons in this regime are intrinsically unstable. This is a consequence of the large group velocity mismatch between the photons exchanged, which gives rise to a Markovian decay process.

Due to the presence of this resonance, the original Hamiltonian $\hat{H}$ is required for the full description of the system dynamics including SH modes in general. Nevertheless, by limiting our interest to the dynamics of a limited band of FH photons and vacuum input to SH, we can still derive a cubic nonlinear Hamiltonian for the FW. The effects of resonant TWM interaction can be taken into account as an effective two-photon loss of FH photons, with the knowledge that the lost population is transferred to the dispersive FH waves. This we accomplish by tracing over the ``extra-band'' dispersive FH modes using standard Born-Markov ME techniques, before introducing the (modified) SW transformation for the ``intra-band'' component. Such a description of the system dynamics is confirmed via numerical simulations. In this case, as the two-photon loss is a higher-order process with respect to the inverse phase-mismatch, we then examine carefully the extent to which the derived equations are perturbatively valid to ensure self-consistency of the theory.

\subsection{Hamiltonian perturbation within momentum band of interest}
The nonlinear term $\hat{V}$ characterizes a process where a SH photon with momentum $p$ is annihilated to generate a pair of FH photons with momentum $\frac{p}{2}\pm q$, and this process is ``resonant'' when $\mathcal{E}_\text{SH}(p)\approx \mathcal{E}_\text{FW}(p,q)$ holds true. Thus, the phase-matching condition forms the curve in the $p$--$q$ plane:
\begin{equation}\label{eq:ellipse}
    \frac{(p-p_0)^2}{a_p^2} + \frac{q^2}{a_q^2} = 1,
\end{equation}
where $p_0 = \gamma/(\frac{1}{2}-\beta)$, $a_p^2 = p_0^2  - 2\xi/(\frac{1}{2}-\beta)$, and $a_q^2 = (\frac{1}{4} - \frac{\beta}{2})a_p^2$. The dispersive coupling regime ($\xi \gg 1$, $\beta < \frac{1}{2}$) corresponds to the case where~\eqref{eq:ellipse} has no solutions. We shall define the dissipative coupling regime to be that where~\eqref{eq:ellipse} has elliptical solutions. Thus, the combination of $p$ and $q$ that fulfills the resonance condition forms an ellipse in the $(p,q)$-plane as shown in Fig.~\ref{fig:dissipative_coupling}(c). This is fulfilled for $\xi < \frac{\gamma^2/2}{1/2-\beta}$, and $\beta < 1/2$. For $\gamma \approx0$, the dissipative coupling regime is thus defined by $\xi \ll -1$.

Based on this knowledge, we decompose $\hat{H}_{\rm NL}$ into an intra-band component
\begin{align}
    \label{eq:vi_integral}
    \hat{V}_\text{i}=\frac{1}{2}\int_{-2p_\text{i}}^{2p_\text{i}}dp\int_{-q_\text{i}(p)}^{q_\text{i}(p)}dq~\left(\hat{\psi}_p^\dagger\hat{\phi}_{\frac{p}{2}+q}\hat{\phi}_{\frac{p}{2}-q}+\mathrm{H.c.}\right)
\end{align}
and extra-band component
\begin{align}
    \hat{V}_\text{e}=\hat{H}_{\rm NL}-\hat{V}_\text{i}
\end{align}
with $q_\text{i}(p)=p_\text{i}-\frac{|p|}{2}$. Here, $p_\text{i}>0$ parametrizes a momentum band of interest $\mathcal{P}_\text{i}=(-p_\text{i},p_\text{i})$, which is chosen such that the region of integration in \eqref{eq:vi_integral} does not contain a resonance as depicted in Fig.~\ref{fig:dissipative_coupling}(c). We derive a reduced quantum model that can properly capture the dynamics of FH photons with momentum within this band, thentrace out SH fields using the mean-field approach.

Since the interaction $\hat{V}_\text{i}$ does not contain a resonance, it can be dealt with perturbatively using the SW transformation. On the other hand, the effects of $\hat{V}_\text{e}$ require
 special care. In fact, as shown in Ref.~\cite{Yanagimoto2020_fano}, such resonant coupling leads to an exponential decay of SH population via a parametric downconversion (PDC) process with characteristic population decay rate of $\pi/\sqrt{|\xi|}$, which we refer to as ``dissipative PDC''. This result can be derived by employing a second-order Born-Markov approach, treating the extra-band component as a disjoint set of ``reservoir'' modes. As the derivation is very similar to those in Appendix~\ref{app:master_equation}, and here we focus on the SW transformation, we relegate this derivation to Appendix~\ref{app:diss_me}.

The result is that we obtain the Lindblad term $\hat{L}_p = \sqrt{\kappa(p)}\hat{\psi}_p$, where the Lindblad contribution to the ME is 
\begin{equation}\label{eq:lindblad_psi}
    \dot{\hat{\rho}} \rightarrow \dot{\hat{\rho}} + \int_{-\infty}^{\infty} dp\left[\hat{L}_p\hat{\rho}\hat{L}^{\dagger}_p - \frac{1}{2}\hat{L}^{\dagger}_p \hat{L}_p \rho - \frac{1}{2}\hat{\rho}\hat{L}^{\dagger}_p \hat{L}_p\right],
\end{equation}
where
\begin{equation}\label{eq:kappa}
\kappa(p) = \frac{\pi}{\sqrt{|\xi|}} + \mathcal{O}(|\xi|^{-3/2}),
\end{equation}
and we also add the Hamiltonian Lamb-shift--like frequency shift term
\begin{equation}\label{eq:G_ls}
    \hat{H}_{\rm LS} = \int_{-\infty}^{\infty} dp \delta(p) \hat{\psi}^{\dagger}_p \hat{\psi}_p,
\end{equation}
where
\begin{equation}\label{eq:delta}
\delta(p) = -\theta q_{\rm i}(p)/|\xi| + \mathcal{O}(|\xi|^{-3/2}).
\end{equation}



\subsection{Schrieffer-Wolff Transformation}
As a result, we now have an effective Hamiltonian without a resonance
\begin{align}
    \hat{H}_{\rm i}=\hat{H}_{\text{i},0}+\hat{V}_\text{i} + \hat{H}_{\rm LS},
\end{align}
where $\hat{H}_{\text{i},0}$ is simply $\hat{H}_{\rm L}$ integrated only over the intra-band momenta, and
with dissipation characterized by \eqref{eq:lindblad_psi}. 
Our next step is to construct a SW transformation $\hat{U}_\text{i}=e^{\hat{S}_i}$ such that 
\begin{align}
    \label{eq:swcondition_dissipative}
    [\hat{H}_{\text{i},0},\hat{S}_\text{i}]=\hat{V}_\text{i}
\end{align}
holds true, with which the Hamiltonian in the interaction frame takes a form
\begin{align}
    \hat{H}'_{\rm i}=\hat{H}_{\text{i},0}+\frac{1}{2}[\hat{S}_\text{i},\hat{V}_\text{i}]+\mathcal{O}(\xi^{-\frac{3}{2}}).
\end{align}
Lindblad operators are also transformed under this basis transformation as
\begin{align}
    \hat{L}'_p= \hat{L}_p+[\hat{S}_\text{i},\hat{L}_p]+\mathcal{O}(\xi^{-\frac{9}{4}}).
\end{align}
Once we obtain the form of $\hat{H}'_\text{i}$ and $\hat{L}'_p$, we drop terms involving dressed field operators for SH (see Apppendix~\ref{app:higherorder} for details).

By following an analogous procedure to the case of dispersive coupling, we find that
\begin{align}
\hat{S}_\text{i}=\int_{-2p_\text{i}}^{2p_\text{i}}dp\int_{-q_\text{i}(p)}^{q_\text{i}(p)}dq~h(p,q)\left(\hat{\psi}_p^\dagger\hat{\phi}_{\frac{p}{2}+q}\hat{\phi}_{\frac{p}{2}-q}-\mathrm{H.c.}\right),
\end{align}
fulfills \eqref{eq:swcondition_dissipative}, where now 
\begin{align}
    \label{eq:fsq}
    h(p,q)
    =-\frac{\theta}{2}\left(\frac{1}{4}(1-2\beta)p^2-\gamma p+\xi+q^2 - \delta(p)\right)^{-1}.
\end{align}
Moving forward, we now approximate the bounds of all integrals over the intra-band region to be $\pm \infty$, as this will not severely affect the dynamics within the band, provided it is sufficiently localized around the center of the band. Under this approximation, it is easy to see that the result for $\hat{H}'_\text{i}$ is similar to that of $\hat{H}'$ for the dispersive coupling regime, but with $f(p,q) \rightarrow h(p,q)$. Thus, dropping terms that involve dressed SH operators, we obtain
\begin{align}\label{eq:spm_diss}
    &\hat{W}_{\text{i,SPM}}=-\frac{1}{4}\iiint dpdqdr~h(p,q)\nonumber\\
    &\quad{}\times\left(\hat{\phi}_{\frac{p}{2}+r}^\dagger\hat{\phi}_{\frac{p}{2}-r}^\dagger\hat{\phi}_{\frac{p}{2}+q}\hat{\phi}_{\frac{p}{2}-q}+\text{H.c.}\right).
    \end{align}
Assuming the bandwidth of excitations to remain limited, we can again in some circumstances approximate $h(p,q)\approx-\theta/2\xi$ (neglecting terms $\mathcal{O}(\xi^{-2})$). In this manner, collecting terms in $\hat{H}_\text{i}'$ that are only composed of FH operators, we obtain an effective Hamiltonian for the dressed FH modes
\begin{align}\label{eq:g77}
    \begin{split}
    \hat{H}_\text{FW}'&=\int_{-\infty}^\infty dp\frac{\theta p^2}{2}\hat{\phi}_p^\dagger\hat{\phi}_p\\
    &+\frac{\theta}{4\xi}\iiint_{-\infty}^\infty dpdqdr~\hat{\phi}_{\frac{p}{2}+q}^\dagger\hat{\phi}_{\frac{p}{2}-q}^\dagger\hat{\phi}_{\frac{p}{2}+r}\hat{\phi}_{\frac{p}{2}-r},
    \end{split}
\end{align}
which takes an identical form as $\hat{H}_\text{cubic}$ up to the difference in basis. The effects of the Lindblad terms \eqref{eq:lindblad_psi} on dressed FH fields are reduced to a form
\begin{align}\label{eq:lfw}
    \hat{L}'_\text{FW,p}=-\theta\sqrt{\frac{\pi}{4}}|\xi|^{-\frac{5}{4}}\int_{-\infty}^\infty dq~\hat{\phi}_{\frac{p}{2}+q}\hat{\phi}_{\frac{p}{2}-q},
\end{align}
where we have also extended the region of integration to infinity.

In Appendix~\ref{app:higherorder}, we show that this theory is valid up to $\mathcal{O}(|\xi|^{-\frac{5}{2}})$.

\subsection{Numerical simulations}
\begin{figure}[hbt]
    \centering
    \includegraphics[width=0.48\textwidth]{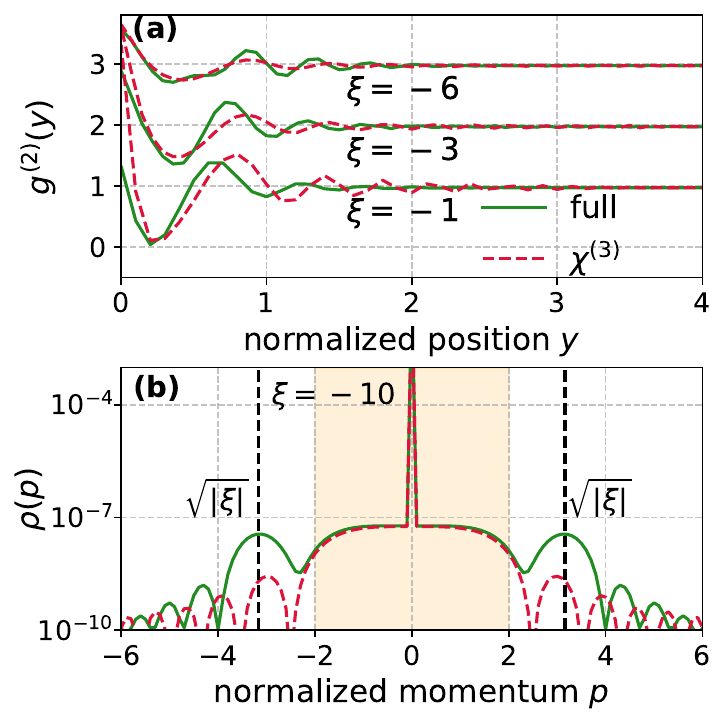}
    \caption{(a) Two-photon spatial correlation function $g^{(2)}(y)$ and (b) momentum distribution $\rho(s)$ of
FH photons prepared initially in a coherent state of $\alpha = 0.1$ for momentum $p = 0$ in the dissipative coupling regime. The green solid lines show the full TWM Hamiltonian (discretized in~\eqref{eq:disc}) result, and the dashed red lines show the naive approximate SW result using the cubic Hamiltonian in Eq.~\eqref{eq:gcubic_discrete}. The orange shaded region shows the intra-band component of the field which we use for the hybrid ME-SW model. Here we let $\gamma = 0$ and $\beta = -1$, and show data at time $t=2$.} 
    \label{fig:attractive}
\end{figure}

First, we simulate the quantum propagation of a weak $p=0$ coherent state in the dissipative coupling regime ($\xi\ll -1$) using the same configurations as Sec.~\ref{sec:dispersive_simulation}. We show the two-photon spatial correlation function $g^{(2)}(y)$ and the momentum distribution $\rho(s)$ for the FH photons simulated using a discretized cascaded quadratic Hamiltonian \eqref{eq:disc} and cubic Hamiltonian \eqref{eq:gcubic_discrete} in Fig.~\ref{fig:attractive}. Notably, unlike the case in the dispersive coupling regime, $g^{(2)}(y)$ of FH under cascaded quadratic nonlinearities exhibits qualitative difference from the corresponding cubic Hamiltonian $\hat{H}_\text{cubic}$. This is due to the high-momentum dispersive FH photons that are generated via the dissipative PDC of SH photons, whose presence can be clearly be seen as peaks around the resonance $p\sim \sqrt{|\xi|}$ in the plot for $\rho(p)$. On the other hand, features within the band of interest away from resonance, which we depict as orange shaded region in the figure, agree well.

\begin{figure}[hbt]
    \centering
    \includegraphics[width=0.48\textwidth]{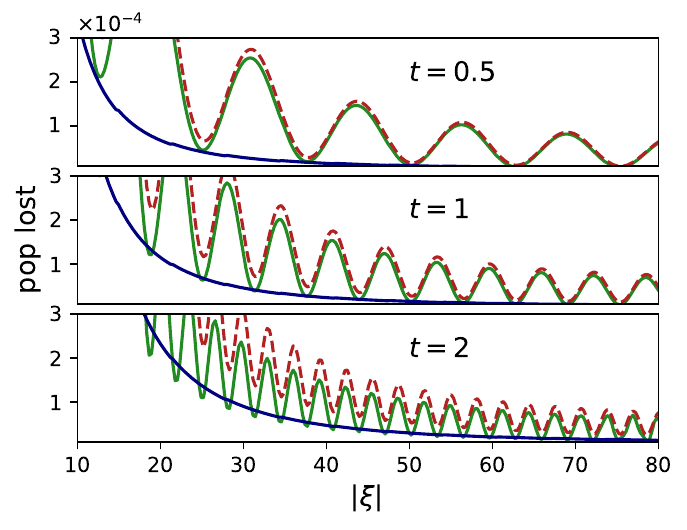}
    \caption{ Population loss $(N_{\rm FW}(t=0)-N_{\rm FW} )/N_{\rm FW}(t=0)$ from an initial lab-frame two-photon Fock state with momentum $p=0$. The full lab-frame Hamiltonian simulation is shown in green, and the SW-transformation cascaded nonlinearity Hamiltonian $\hat{H}'_{\rm FW}$ and Lindblad $\hat{L'}_{\rm FW}$ are shown in dashed red (using the semi-analytic mean-field theory described in Sec.~\ref{sec:mf}). The blue lines show the cascade solution \emph{without accounting} for the SW transformation in either the observable~\eqref{eq:observable} or the initial state. Here we let $\gamma = 0$ and $\beta = -1$.} 
    \label{fig:dissipation}
\end{figure}

Next, to confirm our model in which the dispersive FH component at $p \approx \pm \sqrt{|\xi|}$ is generated by a two-photon loss of FH photons within the intraband component, we introduce the intraband FH population $N_\text{\rm FW}$. When one can neglect the effect of the SW transformation on the definition of this observable, it takes the form
\begin{align}
    N_\text{FW}=\int^{p_\text{i}}_{-p_\text{i}}  dp\hat{\phi}_p^\dagger\hat{\phi}_p.
\end{align}
In Fig.~\ref{fig:dissipation} we plot the photon loss ($(N_\text{FW}(0)-N_\text{FW}(t))/N_\text{FW}(0)$) at time $t$ for the full discretized model~\eqref{eq:disc}, as well as the naive model of the cubic Hamiltonian $\hat{H}'_{\rm FW}$ given in~\eqref{eq:g77} and the two-photon loss Lindblad given in~\eqref{eq:lfw}; we also take these to be discretized in the same manner as before. Clearly, the quantitative as well as qualitative trend in the dissipation is not captured with this naive approach. To rectify this, we must deal with the role that the SW unitary transformation has on the observables corresponding to the FH population, which we do in the next section. Alternatively, we can assume an adiabatic chirping of the phase mismatch $\xi(t)$ to ensure that the SH is initialized in the vacuum state even in the dressed SW frame, which we also discuss in the following section.

 \section{Role of initial conditions in transformation to SW frame: mean-field theory and adiabatic excitation}\label{sec:mf}

In this section, we consider the case where the \emph{lab} frame (as opposed to the SW frame) initially contains no excitations in the SH modes, which is typically the physical condition. In this case, we will show that in the dispersive coupling regime, one can simply neglect the transformation of observables associated with moving from the lab frame to the SW frame and vice-versa, and simply use the cubic cascaded nonlinearity Hamiltonian~\eqref{eq:lieb} as $\xi \rightarrow \infty$. 

In contrast, in the dissipative coupling regime, there are qualitative changes to the dynamics and observables which are associated with the unitary transformation which vanish slower with $|\xi|^{-1}$ than the Lindblad two-photon decay process from the FW. As a result, if we want to accurately describe this higher-order process, we must also self-consistently take into account the role of the transformation. We introduce in this section two methods of doing so: (i) by using a semi-analytic mean-field theory to eliminate the SH degrees of freedom, resulting in an effective theory of only the FH modes, and (ii) by assuming an adiabatic ``chirping'' of the phase-mismatch from a far-detuned value $|\xi|\rightarrow \infty$ to the desired value, which adiabatically populates the dressed eigenstates of the system, which, to the desired order in $|\xi|^{-1}$ are simply the SW frame FH and SH modes.

\subsection{Semi-analytic mean-field theory}

 Here, we assume that we seek a reduced model for the FH dynamics, and thus will treat the dressed SH only to the lowest order in $\xi^{-1}$. We show that the dynamics of the dressed FH can be captured as a mean-field interaction between a free optical meson field (subject to exponential decay in the dissipative coupling case) and the dressed FH subject to the cubic nonlinearity and two-photon decay.

To do this, consider the Hamiltonian~\eqref{eq:lowestperturbation}. From the perspective of the dressed SH subsystem, $\hat{H}_{\rm L} +\hat{W}_{\rm lin}$ is the ``system'' Hamiltonian, and $\hat{W}_{\rm XPM}$ governs the interaction with the dressed FH subsystem, with coupling constant which scales to leading order as $\xi^{-1}$.

Now, one can make a mean-field approximation for the dressed SH by simply tracing out the dressed FH subsystem~\cite{Degenfeld-Schonburg2014Jun}, which simply gives, for the dispersive coupling regime,
\begin{equation}
    \hat{H}'_{\rm SH} = \theta\int^\infty_{-\infty} dp\left(-\xi+\gamma p+\frac{\beta p^2}{2}\right)\hat{\psi}_p^\dagger\hat{\psi}_p + \hat{W}_{\rm lin}+ \mathcal{O}(\xi^{-1}),
\end{equation}
and so $\hat{\psi}_p(t) \approx \hat{\psi}_pe^{-iM(p)t}$---that is, we assume the dressed SH remains as an optical meson state (see Appendix~\ref{app:mesons}). This is similar to the ``undepleted pump approximation'', in which it is assumed that the pump amplitude does not change in magnitude---here, we assume the dressed SH remains close to vacuum. 

In the dissipative coupling regime, we also have the Lindblad term $\hat{L}'_{p,\text{SH}} = \sqrt{\kappa(p)}\hat{\psi}_{p}$. In this case, the Heisenberg operator $\hat{\psi}_p(t)$ takes a slightly more complicated form due to the presence of input noise in the quantum Langevin equations---but nonetheless there exist easily solvable analytical expressions for expectation values of the dressed SH modes (exponential decay, essentially).

It is easy to verify that, under replacing SH operators with their expectation values, interaction terms in the Hamiltonians $\hat{H}'$ and $\hat{H}'_\text{i}$ scale as $\mathcal{O}(|\xi|^{-3})$, and so we can also neglect their influence on the time evolution of the dressed FH modes, just as in the case of initial \emph{dressed} SH in vacuum, as discussed in the previous section. Additionally, there is one cross-channel decay term that arises from the replacement $\hat{L}_p \rightarrow \hat{L}'_p$ under the SW transformation which scales as $\xi^{-\frac{5}{2}}$. This term, after tracing out the dressed SH, manifests as a Hamiltonian term:
\begin{equation}
    \hat{W}_{\rm sq}(t) = i\frac{\theta \pi}{4} |\xi|^{-\frac{3}{2}} \iint_{-\infty}^{\infty} dp dq \langle \hat{\psi}_p\rangle(t) \hat{\phi}_{\frac{p}{2}+q}\hat{\phi}_{\frac{p}{2}-q} + \text{H.c.},
\end{equation}
where $\langle \hat{\psi}_p\rangle(t)$ is given by $\langle \hat{\psi}_p\rangle(t) = \langle \hat{\psi}_p\rangle(t=0)e^{-iM(q)t - \frac{\kappa(q)}{2}t}$. Although the coefficient in front of this term scales as $|\xi|^{-\frac{5}{2}}$, the factor $M(p)$ in the integral oscillates rapidly at frequency $\sim \xi$---thus, the highly detuned oscillations in the dressed FH induced by this squeezing term will have an amplitude $\sim |\xi|^{-\frac{7}{2}}$, and we can neglect this term as well.

Having verified that one can neglect the coupling to the dressed SH entirely when evolving the dressed FH field under the cascade Hamiltonians $\hat{H}'$ and $\hat{H}'_{\text{i}}$, we now turn to the calculation of lab-frame expectation values. In this case, following the mean-field theory, we have the effective initial condition for the \emph{dressed SW frame} reduced density operator for the dressed FH subsystem alone
\begin{align}
    & \text{Tr}_{\psi} \left[ e^{\hat{S}}\ket{\phi_0}\bra{\phi_0} e^{-\hat{S}} \right] = \ket{\phi_0}\bra{\phi_0} \nonumber \\ 
    & + \frac{\sqrt{|\xi|}}{\pi}\mathcal{L}[\hat{L}_{p,\text{FW}}]\ket{\phi_0}\bra{\phi_0} + \mathcal{O}(\xi^{-3}),
\end{align}
where $\ket{\phi_0}$ is the initial density operator for the FH in the lab frame (where we assume the SH is initially in vacuum), $\mathcal{L}[\hat{A}]\hat{\rho} = \hat{A}\hat{\rho}\hat{A}^{\dagger} - \frac{1}{2}\{\hat{A}^{\dagger}\hat{A},\hat{\rho}\}$ is the Lindblad superoperator.

Finally, we note that expectation values can be calculated by applying the SW transformation to their corresponding lab frame operators. For example, the FH photon number can be calculated in the SW frame as 
\begin{align}\label{eq:observable}
   & \langle \hat{\phi}_p^{\dagger} \hat{\phi}_p \rangle_B = \langle e^{\hat{S}}\hat{\phi}_p^{\dagger}\hat{\phi}_p  e^{-\hat{S}}\rangle  \nonumber \\ 
    &=     \langle \hat{\phi}_p^{\dagger} \hat{\phi}_p \rangle  - \frac{\theta}{\xi}\left(\int_{-\infty}^{\infty} \!\!\! dq \langle \hat{\psi}^{\dagger}_{q}\hat{\phi}_p\hat{\phi}_{q-p}\rangle + \text{H.c.}\right) \nonumber \\ 
    & - \frac{1}{4\xi^2}\iint_{-\infty}^{\infty} \!\! dq dq' \left(\langle \hat{\phi}^{\dagger}_p \hat{\phi}^{\dagger}_{q-p} \hat{\phi}_{q-q'}\hat{\phi}_{q'}\rangle + \text{H.c.}\right) + \mathcal{O}(\xi^{-3}).
\end{align}
Here, the expectation value $\langle \hat{\psi}^{\dagger}_q \hat{\phi}_p \hat{\phi}_{q-p}\rangle$ requires some care. This is because the initial state $e^{\hat{S}}\ket{\phi}_0$ contains entanglement between the FH and SH which is, athough small, not negligible to the order of perturbation theory we consider here. Nonetheless, we can use the fact that, in the Heisenberg picture, and within the mean-field approximation,
\begin{equation}
    \hat{\psi}_q(t) = \hat{\psi}_q e^{-i\tilde{M}_q t} + \hat{\zeta}^{(q)}_{\rm in}(t),
\end{equation}
where we let $\tilde{M}_q = M(q) - i\kappa(q)/2$, and
where $\hat{\zeta}^{(q)}_{\rm in}(t)$ represents a vacuum input noise term. This result could be derived using the system-reservoir formalism constructed in Appendix~\ref{app:diss_me}; however, here we simply remark that such a solution to the Heisenberg-Langevin equation is a well-known corollary of the Lindblad form of the loss channels of the SH.

Crucially, $\hat{\zeta}^{(q)}_{\rm in}(t)$ annihilates the total system-environment state vector when the extra-band component begins in vacuum. Thus, using an argument identical to that of the quantum regression theorem for Markovian systems, we can write:
\begin{align}
\langle \hat{\psi}^{\dagger}_{q}\hat{\phi}_p\hat{\phi}_{q-p}\rangle & = \text{Tr}\left[\hat{\rho}_{\rm S}(0)\hat{\rho}_{\rm R}(0)\hat{\psi}_q^{\dagger}(t)\hat{\phi}_p(t)\hat{\phi}_{q-p}(t) \right] \nonumber \\ 
& = e^{iM^*_q t}\text{Tr}\left[\hat{\phi}_{p}\hat{\phi}_{q-p}e^{\mathbb{L}t}\hat{\rho}_S(0)\hat{\psi}_q^{\dagger}\right],
\end{align}
where $\hat{\rho}_{\rm S}$ denotes the system (intraband FH and SH), $\hat{\rho}_{\rm R}$ denotes the reservoir (extraband FW) (see Appendix~\ref{app:diss_me} for details), and $\mathbb{L}$ is the Liouvillian of the system.
Now, since $\hat{\rho}_{\rm S}(0) = e^{\hat{S}}\ket{\phi_0}\bra{\phi_0}e^{-\hat{S}}$, we have
\begin{equation}
    \hat{\rho}_S(0)\hat{\psi}_q^{\dagger} = \int_{-\infty}^{\infty} \!\! dq' h(q,q') \ket{\phi_0}\bra{\phi_0}\hat{\phi}^{\dagger}_{\frac{q}{2} + q'}\hat{\phi}^{\dagger}_{\frac{q}{2}-q'} + \mathcal{O}(\xi^{-2}),
\end{equation}
and thus, we can write
\begin{align}
\langle \hat{\psi}^{\dagger}_{q}\hat{\phi}_p\hat{\phi}_{q-p}\rangle & \approx \int_{-\infty}^{\infty} dq' h(q,q') e^{i\tilde{M}^*_q t} \nonumber \\ & \times \text{Tr}\left[\hat{\phi}_p \hat{\phi}_{q-p} e^{\mathbb{L}t}\left(\ket{\phi_0}\bra{\phi_0}\hat{\phi}^{\dagger}_{\frac{q}{2}+q'}\hat{\phi}^{\dagger}_{\frac{q}{2}-q'}\right)\right],
\end{align}
and the trace can be straightforwardly evaluated as an expectation value $\langle \hat{\phi}_p \hat{\phi}_{q-p}\rangle$ under the \emph{effective} density operator $\chi_{qq'}(t)$, where $\chi_{qq'}(t=0) = \ket{\phi_0}\bra{\phi_0}\hat{\phi}^{\dagger}_{\frac{q}{2}+q'}\hat{\phi}^{\dagger}_{\frac{q}{2}-q'}$, and $\chi_{qq'}$ follows the same ME as that of the system density operator in the SW frame under the mean-field approximation. Thus, any requirement of numerically simulating the SH is removed from the analysis.  

However, in our case, the dynamics of the FH to leading order can be analytically captured by using only the dispersion Hamiltonian $\hat{H}_{\text{i},0}$. Thus we can further approximate
\begin{align}
\langle \hat{\psi}^{\dagger}_{q}\hat{\phi}_p\hat{\phi}_{q-p}\rangle & \approx \int_{-\infty}^{\infty} dq' h(q,q') e^{iM^*_q t} \nonumber \\ & \times e^{-i\frac{p^2+(q-p)^2}{2}t}\langle \hat{\phi}^{\dagger}_{\frac{q}{2}+q'}\hat{\phi}^{\dagger}_{\frac{q}{2}-q'} \hat{\phi}_p \hat{\phi}_{q-p}\rangle_0 \nonumber \\ 
& \approx -\frac{\theta}{2\xi}e^{-i\left(\xi + \frac{\pi}{2\sqrt{|\xi}}\right)t}\int_{-\infty}^{\infty} \!\!\! dl \langle \hat{\phi}^{\dagger}_l \hat{\phi}^{\dagger}_{q-l} \hat{\phi}_p \hat{\phi}_{q-p}\rangle_0,
\end{align}
where in the second line we have further approximated the dispersion relation and decay rates to leading order, and we let the nought subscript denote the value at $t=0$ in the dressed frame. 

We remark that since the SW transformation induces only a perturbative few-photon excitation in the dressed frame initial condition, we can always follow this process of analytically removing the initial SH-SW entanglement in favor of a two-time correlation of the FH variables by means of the procedure outlined in this section. Equivalently, we could cast the above heuristic derivation in terms of a formal Feshbach projection approach, where we solve for zero- and few-SH--photon subsystems simultaneously.

In Fig.~\ref{fig:dissipation}, we show in red dashed lines the FH photon loss following the semi-analytic mean-field theory outlined in the section. Clearly, the loss dynamics approach the full result in the limit $|\xi| \rightarrow \infty$; at longer times, one must go to larger and larger $|\xi|$ to see sufficient agreement.
\subsection{Adiabatic excitation of dressed FH modes}\label{sec:adiabatic}

One approach to simplify the modelling of the cascaded nonlinearity is to assume that one can excite the dressed FH state by means of an adiabatic turn-on and turn-off of the interaction Hamiltonian $\hat{H}_{\rm NL}$. This can be done by modulating the phase-mismatch $\xi \rightarrow \xi(t)$. Physically, this parameter change corresponds to a chirping in the poling
period of a quasi-phase matched device, for instance.

Specifically, consider the lab frame full Hamiltonian $\hat{H}$ at time $t = 0$ with a very large phase-mismatch, such that 
\begin{equation}
\hat{H}(t=0)  \approx -\xi_i \int_{-\infty}^{\infty} dp \hat{\psi}_p^{\dagger}\hat{\psi}_p,
\end{equation}
where $|\xi_i| \ggg 1$---that is, sufficiently large as to be able to consider SH photon number states as eigenstates of the system with photon energy $-\xi_i$. Then, consider a smoothly varying $\xi(t)$ such that at time $t_1$, $\xi(t_1) = \xi_f$, where $|\xi_f| \ll |\xi_i|$.

By moving to the SW frame, it can be seen from the discussion in Appendix~\ref{app:higherorder} that the eigenstates of $\hat{H}(t_0)$ are, to order $|\xi|^{-5}$, simply the FH and SH excitations (in the dressed frame). Thus, if we start in an eigenstate (lab frame FH excitation) at time $t= 0$, we can adiabatically excite the dressed FH states by means of varying $\xi(t)$ sufficiently slowly. Similarly, the dressed states can be converted back into physical FH excitations by an adiabatic increase of the phase-mismatch magnitude after the interaction with the nonlinear crystal. In Appendix~\ref{app:single} we show that this procedure can effectively remove the need to employ any transformations in the analysis; one can simply use the cascaded Hamiltonian for straightforward calculations.

\section{Conclusions}\label{sec:conclusions}

In conclusion, we have presented EFT techniques for the reduction of the band degrees of freedom of quantum nonlinear optical systems in perturbative regimes. These techniques motivate an emergent system-reservoir separation, and can handle both dissipative and dispersive processes in a rigorous and systematic fashion. 

We first outlined the general approach and introduced a geometric picture of interaction regimes in the ``interaction space'' determined by the number of photons involved in the coupling Hamiltonian, which can allow one to determine where dissipative and dispersive behaviour will occur. After describing the general approach, we focused on a case study of the techniques to the application of propagation through a $\chi^{(2)}$-nonlinear waveguide where the center of the FH and SH bands have a large phase-mismatch. We showed that the functional form of the dispersion of the energy bands determined the form of the interaction regime, identifying the textbook dispersive coupling regime as well as a dissipative coupling regime. The dispersive coupling regime recovered the expected effective $\chi^{(3)}$ cascaded nonlinear Hamiltonian, while the dissipative coupling regime also contains a dissipative loss channel from the SH to dispersive FH waves, which we derived using master equation techniques. We confirmed that under the SW transformation, this loss channel can manifest as effective two-photon loss from the center of the FH band. This higher-order effect requires an analysis beyond standard leading-order perturbation theory, which we were able to quantitatively incorporate through a semi-analytic mean-field theory, taking advantage of the few-photon subsystem of excitation of the SH within the SW transformation. Beyond cascaded $\chi^{(2}$, we expect this framework to be of immediate application to $\chi^{(3)}$ OPA-type systems, as well as the quantum modelling of parasitic high-harmonic generation in broadband nonlinear platforms.

This analysis shows the utility and power of our approach, which we expect to be a useful modelling tool for simulations and gaining physical intuition about ultrafast nonlinear optical systems. Our fully quantum approach is particularly relevant as recent experimental progress in single-photon nonlinearities drives forward towards the regime where quantum and non-Gaussian effects become important, and simultaneous advances in dispersion engineering open up a broad vision of many-body physics and Hamiltonian engineering using interacting photons as a platform.

\acknowledgments
We thank Noah Fleming for useful discussions. 
\begin{appendix}
\section{Nondimensionalization of $\chi^{(2)}$ Hamiltonian}\label{app:nondimensionalization}


This appendix gives extra detail on the scaling procedure to obtain the dimensionless Hamiltonian used in Eq.~\eqref{eq:G_Hamiltonian}. We start with the Hamiltonian (in physical units)
\begin{align}
\hat{\tilde{H}}=\hat{\tilde{H}}_\mathrm{L}+\hat{\tilde{H}}_\mathrm{NL}
\end{align}
with linear part
\begin{align}
\hat{\tilde{H}}_\mathrm{L}=\sum_{u\in \{a,b\}}\int dz\,\hat{u}_z^\dagger \delta\omega_u(i\partial_z)\hat{u}_z
\end{align}
and nonlinear part
\begin{align}
\hat{\tilde{H}}_\mathrm{NL}=\frac{r}{2}\int dz\,(\hat{a}_z^{\dagger 2}\hat{b}_z+\hat{a}_z^{2}\hat{b}_z^\dagger),
\end{align}
where $r$ is the $\chi^{(2)}$ nonlinear coupling strength, and $\hat{a}_z$ and $\hat{b}_z$ are spatial FH and SH annihilation operators with $[\hat{a}_z,\hat{a}_{z'}^\dagger]=[\hat{b}_z,\hat{b}_{z'}^\dagger]=\delta(z-z')$. In a rotating frame, the ``residual'' dispersion around the carriers $\delta\omega_u(k)$ ($u=a$ for FH and $u=b$ for SH) are given in terms of the waveguide dispersion $\omega(k)$ as
\begin{subequations}
\begin{align}
\delta\omega_a(k)&=\omega(k+k_0)-\omega_\mathrm{ref}-kv_\mathrm{ref}\\
\delta\omega_b(k)&=\omega(k+2k_0)-2\omega_\mathrm{ref}-kv_\mathrm{ref},
\end{align}
\end{subequations}
where $k_0$ is the FH carrier angular-wavevector, $\omega_\mathrm{ref}$ and $v_\mathrm{ref}$ are reference phase- and group-velocity, respectively. Note that the spatial coordinate $z$ is defined in the co-propagating frame with $v_\mathrm{ref}$. While the choice of $\omega_\mathrm{ref}$ and $v_\mathrm{ref}$ is arbitrary, for later convenience, we choose them so that $0$th and $1$st order terms of the signal dispersion is eliminated, i.e., $\omega_\mathrm{ref}=\omega(k_0)$ and $v_\mathrm{ref}=\omega'(k_0)$. Expanding dispersions up to second order, we obtain
\begin{subequations}
\begin{align}
\delta\omega_a(k)&=\frac{1}{2}\omega_a''(k_0)k^2 \\
\delta\omega_b(k)&=\delta\omega_b(0)+\delta\omega_b'(0)k+\frac{1}{2}\omega_b''(2k_0)k^2
\end{align}
\end{subequations}

Then, we move to a wavespace via Fourier transform
$\hat{u}_s=\int dz\,e^{-2\pi i sz}\hat{u}_z$,
where the wavenumber $s$ is related to the angular wavenumber $k$ via $k=2\pi s$. 

We then nondimensionalize the Hamiltonian, by introducing the characteristic frequency
\begin{align}
    g_\mathrm{c}=\left(\frac{r^4}{4\pi^2|\omega''(k_0)|}\right)^{1/3}
\end{align}
and length
\begin{align}
z_\mathrm{c}=\left(\frac{4\pi^2|\omega''(k_0)|}{r}\right)^{2/3},
\end{align}
with which we normalize the Hamiltonian (i.e., time) and wavenumber as
\begin{align}
    &\hat{H}=\hat{H}/g_\mathrm{c} &p=sz_\mathrm{c}.
\end{align}
The field operators are also scaled as
\begin{align}
&\hat{\phi}_p=[z_\mathrm{c}]^{-\frac{1}{2}}\,\hat{a}_s, &\hat{\psi}_p=[z_\mathrm{c}]^{-\frac{1}{2}}\,\hat{b}_s.
\end{align}
Thus, we obtain the result found in Eq.~\eqref{eq:G_Hamiltonian}. $\theta=\mathrm{sign}(\omega''(k_0))$ is the sign of signal group-velocity dispersion, $\xi=-\theta\delta\omega_b(0)/g_\mathrm{c}$ is a normalized phase-mismatch, $\gamma=\theta2\pi \omega_b'(0)/g_\mathrm{c}z_\mathrm{c}$ is a normalized group-velocity mismatch, and $\beta=\omega''(2k_0)/|\omega''(k_0)|$. 

\section{Perturbative validity of SW expansion}\label{app:higherorder}
\subsection{Dispersive coupling regime}
Performing the SW expansion to the next higher-order term (after what was derived in Sec.~\ref{sec:dispersive_coupling}) gives
\begin{equation}
    \hat{H}' = \hat{H}_{\rm L} + \hat{W}_{\rm SPM} + \hat{W}_{\rm lin} + \hat{W}_{\rm XPM} + \left(\hat{V}' + \text{H.c.}\right),
\end{equation}
where,
\begin{widetext}
\begin{align}\label{eq:SWnextorder}
&\hat{V}' = \frac{\theta \pi}{2} \iint_{-\infty}^{\infty} dp dq \left[\frac{f(p,q)}{\sqrt{\omega(p)}} - \frac{\theta}{12\left[\omega(p)\right]^{\frac{3}{2}}}\right]\hat{\psi}_p^{\dagger}\hat{\phi}_{\frac{p}{2}+q}\hat{\phi}_{\frac{p}{2} -q}  \nonumber \\ 
& +\frac{4}{3} \!\iiiint_{-\infty}^{\infty} \! \! dp dq dp' dq' f(p,q)\hat{\psi}^{\dagger}_p \left[f(p',q')\hat{\psi}^{\dagger}_{p'}\hat{\psi}_{\frac{p+p'}{2} + q +q'} + f\left(p',\frac{p'-p}{2}-q\right)\hat{\psi}^{\dagger}_{\frac{3p'-p}{2}- q -q'} \hat{\psi}_{p'}\right]\hat{\phi}_{\frac{p'}{2}-q'}\hat{\phi}_{\frac{p}{2} -q} \nonumber \\
& \!-\!\frac{2}{3} \!\iiiint_{-\infty}^{\infty} \!\!\!\!\!\! dp dq dp' dq' \hat{\psi}^{\dagger}_{p'} \!\left[f(p,q)f(p,q')\hat{\phi}^{\dagger}_{\frac{p}{2}\!-\!q'}\hat{\phi}_{p'\!-\!\frac{p}{2}\!-\!q'} + f(p',q')\!\left(\!2f(p,q) + f(p,\frac{p'\!-\!p}{2}\!+\!q')\right)\hat{\phi}^{\dagger}_{p\!-\!\frac{p'}{2}\!-\!q'}\hat{\phi}_{\frac{p'}{2}\!-\!q'} \right] \hat{\phi}_{\frac{p}{2} +q}\hat{\phi}_{\frac{p}{2}-q}  
\end{align}
\end{widetext}
Despite the complexity of~\eqref{eq:SWnextorder}, we note that all the terms either excite one net dressed SH photon and annihilate two net dressed FH photons, or vice-versa. Moreover, the top line scales as, to leading order, $\xi^{-\frac{3}{2}}$, and the rest scale as $\xi^{-2}$. Thus, as long as the number of photons in the system is not excessively large, we can focus on the first line to determine the scaling of excitations in the SW frame in the large phase-mismatched limit $\xi \gg 1$.

Neglecting all but the top line, we can use similar heuristic arguments to those in Sec.~\ref{sec:classical} to deduce that excitations induced in the dressed SH modes will scale as $\hat{\psi} \sim \xi^{-\frac{5}{2}}$. This can also be seen by considering perturbation theory with $\hat{V}' + \text{H.c.}$ as the perturbation Hamiltonian, and using the dressed FH photon Fock states as the unperturbed eigenstates in the SW frame (in this case, revealing that the Fock states are hybridized with SH excitations to leading order $\xi^{-5}$). As a result, the SH excitations induced by the FH can be safely neglected, as discussed in Sec.~\ref{sec:dispersive_coupling}.

The next-to-leading term in the SW expansion that does not depend on the SH operators scales as $\xi^{-\frac{5}{2}}$, and modifies the SPM interaction. We can incorporate this term by by changing $\hat{W}_{\rm SPM}$ to $\hat{W}'_{\rm SPM}$, where
\begin{align}\label{eq:Wprime}
    \hat{W}'_\text{SPM}=&-\!\frac{1}{4}\iiint^\infty_{-\infty} \!\!\!\!\!  dp dq drf'(p,q,r)\hat{\phi}_{\frac{p}{2}+r}^\dagger\hat{\phi}_{\frac{p}{2}-r}^\dagger\hat{\phi}_{\frac{p}{2}+q}\hat{\phi}_{\frac{p}{2}-q}\nonumber\\
    &\quad{}+\mathrm{H.c.},
\end{align}
and $f(p,q,s) = f(p,q)\left[1+l(p,r)\right]$, where
\begin{equation}
    l(p,r) = \theta\frac{3 \pi}{4}\frac{\left[f(p,r) - \theta/(12\omega(p))\right]}{\sqrt{\omega(p)}}.
\end{equation}
The replacement $\hat{W}_{\rm SPM} \rightarrow \hat{W}'_{\rm SPM}$ changes the validity of the SW expansion Hamiltonian $\hat{H}'$ from $\mathcal{O}(\xi^{-\frac{5}{2}})$ to $\mathcal{O}(\xi^{-3})$. 

\subsection{Dissipative coupling regime}
In the dissipative coupling regime, the Hamiltonian containing the leading-order terms that couple to the dressed SH modes is
\begin{equation}
    \hat{H}'_{\rm i} = \hat{H}_{\text{i},0} + \hat{H}_{\rm LS} + \hat{W}_{\text{i},\text{lin}} + \hat{W}_{\text{i},\text{SPM}} + \hat{W}_{\text{i},\text{XPM}} + \mathcal{O}(|\xi|^{-\frac{3}{2}}),
\end{equation}
where $\hat{W}_{\text{i},\text{SPM}}$ is given in~\eqref{eq:spm_diss}, and 
\begin{align}
\hat{W}_\text{i,XPM}=&\iiint_{-\infty}^{\infty}   d p  dq drh(p,q)\hat{\psi}_p^\dagger\hat{\psi}_r\hat{\phi}^\dagger_{r-\frac{p}{2}-q}\hat{\phi}_{\frac{p}{2}-q}\nonumber\\
    &+\mathrm{H.c.},
\end{align}
and
\begin{align}
    \hat{W}_\text{i,lin}=\frac{\theta}{2}\int^\infty_{-\infty}\frac{ dp}{\sqrt{|\omega'(p)|}}\text{arctanh}\left(\frac{q_{\rm i}(p)}{\sqrt{|\omega'(p)|}}\right)\hat{\psi}_p^\dagger\hat{\psi}_p,
\end{align}
where 
\begin{align}
    \label{eq:omegap}
   \omega'(p)=\frac{p^2}{4}-\theta\mathcal{E}_\text{SH}(p) - \theta \delta(p).
\end{align}

It is straightforward to see that the next term in the SW for the dissipative coupling regime is similar in form to that of the dispersive regime ~\eqref{eq:SWnextorder}, but with $f(p,q) \rightarrow h(p,q)$ and $\omega(p) \rightarrow \omega'(p)$. As a consequence, the arguments in the previous section regarding the excitations induced in the dressed SH modes due to interactions with the dressed FH are very similar in the dissipative coupling case. 

However, by assumption,  $q_{\rm i}(p) \ll \sqrt{|\omega'(p)|} \sim |\xi|^{\frac{1}{2}}$, and so to leading order, $\hat{W}_{\rm i, lin}$ scales as $\sim |\xi|^{-1}$, in contrast to the $\sim |\xi|^{-\frac{1}{2}}$ scaling in the dispersive case. Following this argument through the calculations in the previous subsection, we deduce that the SPM Hamiltonian $\hat{W}_{\rm i, SPM}$ is in fact valid to $\mathcal{O}(|\xi|^{-3})$ without any additional corrections, in contrast to the dispersive coupling regime. 

One more difference is that with the transformation $\hat{L}_p \rightarrow \hat{L}'_p$, additional cross-channel decay terms appear in the ME in the SW frame, which are not in Lindblad form. To this end, let us consider $\hat{L}'_p$ up to the $|\xi|^{-\frac{9}{4}}$ term (third term in SW transformation expansion):
\begin{align}
   & \frac{\hat{L}'_p}{\sqrt{\kappa(p)}} = \frac{\hat{L}_p}{\sqrt{\kappa(p)}}  -\int_{-\infty}^{\infty} dq h(p,q) \hat{\phi}_{\frac{p}{2}+q}\hat{\phi}_{\frac{p}{2}-q} \nonumber \\
   & - 2 \iint_{-\infty}^{\infty} dp' dq h(p,q)h(p',\frac{p-p'}{2}+q)\hat{\psi}_{p'}\hat{\phi}^{\dagger}_{p'-\frac{p}{2}-q}\hat{\phi}_{\frac{p}{2}-q} \nonumber \\ 
    &-\frac{1}{4|\omega'(p)|^{\frac{3}{2}}}\text{arctanh}\left(\frac{q_{\rm i}(p)}{\sqrt{|\omega'(p)|}}\right)\hat{\psi}_p  + \mathcal{O}(|\xi|^{-3}).
\end{align}

Importantly, none of the cross-channel decay terms can create photons in the dressed SH. Moreover, taking into account the expected $\hat{\psi} \sim |\xi|^{-\frac{5}{2}}$ scaling, they are negligible for the sake of the Hamiltonian dynamics when the dressed SH starts off in vacuum.

\section{Optical Mesons}\label{app:mesons}
In this appendix we provide further insights on the physical interpretation of the dressed SH modes. While the linear energy dispersion of the dressed FH remains the same in the interaction frame, dressed SH modes experience energy shifts due to $\hat{W}_\text{lin}$. The total linear energy dispersion of $M(p)$ becomes
\begin{align}
    \label{eq:mesondispersion}
&M(p)=\mathcal{E}_\text{SH}-\frac{\pi\theta}{2}\left(\frac{p^2}{4}-\theta\mathcal{E}_\text{SH}\right)^{-\frac{1}{2}}\\
&=\theta \left(-\xi+\gamma p+\frac{\beta p^2}{2}-\frac{\pi}{2}\left\{\frac{(1-2\beta)p^2}{4}-\gamma p+\xi\right\}^{-\frac{1}{2}}\right).\nonumber
\end{align} 
Here, the dressed SH excitation in the interaction frame corresponds to an excitation represented by an annihilation
\begin{align}
    &\hat{\mu}_p=\hat{\psi}_p+[-\hat{S},\hat{\psi}_p]+\mathcal{O}(\hat{S}^2)\\
    &=\hat{\psi}_p+\int^\infty_{-\infty} dqf\left(p,r\right)\hat{\phi}_{\frac{p}{2}+q}\hat{\phi}_{\frac{p}{2}-q}+\mathcal{O}(\xi^{-2})\nonumber
\end{align}
in the lab frame, whose linear energy dispersion is given as $M(p)$. Interestingly, $\ket{\mu_p}=\hat{\mu}_p^\dagger\ket{0}$ turns out to be approximately optical mesons~\cite{Drummond1997}, which are dressed single-SH-photon eigenstates of \eqref{eq:G_Hamiltonian}. We review the derivations of optical mesons to show their correspondence to the dressed SH modes below.

We posit the form of an eigenstate of \eqref{eq:G_Hamiltonian} as
\begin{align}
    \varphi_p^\dagger\ket{0}=\left(\hat{\psi}_p^\dagger+\int^{\infty}_0 dq g(p,q)\hat{\phi}_{\frac{p}{2}+q}^\dagger\hat{\phi}_{\frac{p}{2}-q}^\dagger\right)\ket{0}
\end{align}
with an eigenenergy $\mathcal{M}(p)=-\theta \mathcal{E}_p$. As $\hat{H}\ket{\varphi_p}=\mathcal{M}(p)\ket{\varphi_p}$ has to be fulfilled for $\ket{\varphi_p}=\varphi_p^\dagger\ket{0}$, we obtain
\begin{subequations}
\begin{align}
    \label{eq:sub1}
    \theta \left(-\xi+\gamma p+\frac{\beta p^2}{2}\right)+\frac{1}{2}\int^\infty_{-\infty} dq g(p,q)=-\theta \mathcal{E}_p\\
    \label{eq:sub2}
    \frac{1}{2}+\theta\left(q^2+\frac{p^2}{4}\right)g(p,q)=-\theta \mathcal{E}_pg(p,q).
\end{align}
\end{subequations}
By solving \eqref{eq:sub2} assuming $\mathcal{E}_p>0$, we have
\begin{align}
    \label{eq:gfunc}
    g(p,q)=-\frac{\theta}{2\left(q^2+\frac{p^2}{4}+\mathcal{E}_p\right)},
\end{align}
and substituting \eqref{eq:gfunc} into \eqref{eq:sub1} leads to
\begin{align}
    \label{eq:mesonenergy}
    \mathcal{E}_p-\frac{\pi}{2\sqrt{\mathcal{E}_p+\frac{p^2}{4}}}=\xi-\gamma p-\frac{\beta p^2}{2}.
\end{align}
This equation can be solved perturbatively to yield
\begin{align}
    \mathcal{M}(p)=M(p)+\mathcal{O}(\xi^{-2}).
\end{align}
As a result, we have 
\begin{align}
    g(p,q)&=-\frac{\theta}{2\left(\frac{1}{4}(1-2\beta)p^2-\gamma p+\xi+q^2+\mathcal{O}(\xi^{-\frac{1}{2}})\right)}\nonumber\\
&=f(p,q)+\mathcal{O}(\xi^{-\frac{5}{2}}),
\end{align}
resulting in
\begin{align}
    \ket{\mu_p}=\ket{\varphi_p}+\mathcal{O}(\xi^{-\frac{5}{2}}).
\end{align}
These results show that the dressed pump excitations that are annihilated by $\hat{\psi}_p$ in the interaction frame approximately correspond to optical mesons in the lab frame.

%

\section{ME approach to cascaded $\chi{(2)}$ nonlinearity}\label{app:master_equation}

In Sec.~\ref{sec:sw}, we outlined a SW approach to treating the cascaded $\chi^{(2)}$ nonlinearity in the dispersive and dissipative coupling regimes, by constructing a perturbative unitary transformation to remove the dispersive TWM interaction and diagonalize the Hamiltonian. In the dissipative coupling regime, we used a ME only to remove an ``extra-band'' component of the FH modes which was involved in a  decay channel from the SH.

Alternatively, in this section we present a theory of multimode cascaded nonlinearity using exclusively a ME theory. This approach, also valid in the limit $|\xi| \gg 1$, has the advantage of not requiring a unitary frame transformation, and explicitly eliminates the SH degree of freedom. We show that the standard results regarding the derivation of the cubic Hamiltonian can also be derived using this method. In contrast to the hybrid SW transformation-ME approach of Sec.~\ref{sec:sw}, this approach does not capture the effective cross-phase modulation interaction described in Sec.~\ref{sec:dispersive_coupling}, nor does it capture the higher-order effect outlined in Fig.~\ref{fig:dissipative_coupling} of two-photon loss from the ``intra-band'' FH modes to the extra-band modes, mediated by the dispersive interaction with the SH.  In the dissipative regime, this approach can capture the resonant process of FH modes with momentum $\sim \pm \sqrt{|\xi|}$ interacting with the SH with momentum $\sim 0$, but in this case gives the reverse spontaneous decay process from the FH to the SH.

To derive the ME, we use a standard second-order time-convolutionless Born-Markov approach~\cite{breuer2002}, treating the FH as the system and the SH as a reservoir. Here, we can treat both dispersive and dissipative coupling regimes simultaneously. 

We  separate the total Hamiltonian $\hat{H} = \hat{H}_{\rm L} + \hat{H}_{\rm NL}$ into system, reservoir, and interaction parts, such that $\hat{H} = \hat{H}_{\rm S} + \hat{H}_{\rm R} + \hat{H}_{\rm NL}$, where $\hat{H}_S + \hat{H}_R = \hat{H}_{\rm L}$, such that
\begin{equation}
    \hat{H}_{\rm S} = \int_{-\infty}^{\infty} dp \frac{p^2}{2}\hat{\phi}_p^{\dagger}\hat{\phi}_p
\end{equation}
and
\begin{equation}
\hat{H}_{\rm R} = \int_{-\infty}^{\infty} dp \left[-\xi + \gamma p + \frac{\beta p^2}{2}\right]\hat{\psi}^{\dagger}_p \hat{\psi}_p
\end{equation}
are the system and reservoir Hamiltonians, respectively. The total density operator for the combined system + reservoir $\hat{\rho}$ then follows the von Neumann equation
\begin{equation}\label{eq:vn}
    \frac{ d}{d t}\hat{\tilde{\rho}}(t) = -i[\hat{\tilde{G}}_{\rm NL}(t),\hat{\tilde{\rho}}(t)],
\end{equation}
where we have moved into the interaction picture by means of $\hat{\tilde{\rho}}(t) = e^{i\hat{H}_{\rm L}t}\hat{\rho}(t)e^{-i\hat{H}_{\rm L}t}$.
Integrating~\eqref{eq:vn} then substituting the result back into~\eqref{eq:vn} and tracing over the reservoir yields
\begin{align}\label{eq:vn2}
    \frac{ d}{d t}\hat{\tilde{\rho}}_{\rm S}(t) = \int_0^{t-t_0} \!\! \!\! d\tau \text{Tr}_{\rm R}\left[\hat{\tilde{G}}_{\rm NL}(t\!-\!\tau)\hat{\tilde{\rho}}(t\!-\!\tau),\hat{\tilde{G}}_{\rm NL}(t)\right] + \text{H.c.}
\end{align}
where $\hat{\tilde{\rho}}_{\rm S}$ is the reduced system density operator for the FW. Here, we have assumed $\text{Tr}_{\rm R}[\hat{\tilde{G}}_{\rm NL}(t),\hat{\tilde{\rho}}(t_0)] =0$, which is satisfied provided the system is in a separable state at $t =t_0$ wherein the SH has vanishing mean amplitude. Here we will assume this corresponds to vacuum initial condition for the SH, but it is also straightforward to consider thermal states, as well as coherent and squeezed states (by means of unitary transformation).

Next, we eliminate the SH degrees of freedom by making the conventional second-order Born-Markov approximation, which involves replacing $\hat{\tilde{\rho}}(t-\tau)$ under the integral with $\hat{\tilde{\rho}}_{\rm S}(t)\hat{\tilde{\rho}}_{\rm R}(t_0)$. We give intuitive arguments for the validity of this approximation in the highly phase-mismatched $|\xi| \gg 1$ regime in Appendix~\ref{app:diss_me}, and note that this heuristic replacement can also be derived using formal projection operator techniques, which also allows one to systematically derive higher-order corrections~\cite{breuer2002}.

Finally, we extend the upper limit of the integral in~\eqref{eq:vn2} to $+\infty$; we assume at time $t_0=0$, the SH modes are in vacuum (a product state with the FW) and the nonlinearity is suddenly switched on. This replacement is then called the ``second Markov approximation'', which is justified in the same regime of validity as the Born-Markov approximation.

 Moving back into the Schr{\"o}dinger picture, we thus have the Born-Markov ME in general form:
 \begin{equation}
     \frac{ d}{d t}\hat{\rho}_{\rm S}(t) = -i[\hat{H}_{\rm S},\hat{\rho}_{\rm S}(t)] + \mathbb{L}\hat{\rho}_\text{S},
 \end{equation}
 where the ME contribution from the reservoir interaction is 
\begin{equation}\label{eq:me_dissipator}
\mathbb{L}\hat{\rho}_{\rm S} = \int_0^{\infty} \!\!\! d\tau \text{Tr}_{\rm R}\Big[\hat{\tilde{G}}_{\rm NL}(-\tau)\hat{\rho}_{\rm S}(t)\hat{\rho}_{\rm R}(t_0),\hat{H}_{\rm NL}\Big] + \text{H.c.},
\end{equation}
where we have neglected higher-order terms beyond the second-order Born-Markov approximation, which we presume to scale as $\mathcal{O}(|\xi|^{-2})$.

Noting that we can write
\begin{equation}
    \hat{\tilde{G}}_{\rm NL}(-\tau) = \frac{1}{2} \iint_{-\infty}^{\infty}\!\!\!  dp dq \hat{\psi}_p \hat{\phi}^{\dagger}_{\frac{p}{2}+q}\hat{\phi}^{\dagger}_{\frac{p}{2}-q} e^{\frac{i}{2}\left[f(p,q)\right]^{-1}\tau} + \text{H.c.},
\end{equation}
it is straightforward to apply the partial trace and evaluate~\eqref{eq:me_dissipator}. Assuming vacuum for the SH, and applying the Sokhotski–Plemelj theorem $\int_{\infty}^{\infty} dt e^{i\omega t} = \pi \delta(\omega) + i\text{P}(1/\omega)$, we find
\begin{align}
    \mathbb{L}& \hat{\rho}_{\rm S}(t) = \iiint_{-\infty}^{\infty} \!  dp dq dq' \zeta(p,q') \nonumber \\ & \times \left[\hat{\phi}_{\frac{p}{2} + q'}\hat{\phi}_{\frac{p}{2}-q'}\hat{\rho}_{\rm S}, \hat{\phi}^{\dagger}_{\frac{p}{2}+q}\hat{\phi}^{\dagger}_{\frac{p}{2}-q}\right] + \text{H.c.},
\end{align}
where
\begin{equation}\label{eq:zeta}
    \zeta(p,q') = \frac{\pi}{4}q_0^{-1}(p)\delta(q'-q_0(p)) - \frac{i}{2}\text{P}f(p,q')
\end{equation}
where $q_0(p) =\sqrt{-\xi + \gamma p + \frac{1}{4}(2\beta-1)p^2}$ is the real positive solution in terms of $q'$ to $\left[f(p,q')\right]^{-1}= 0$; in the dispersive regime with $\xi \gg 1$ and $\beta < -\frac{1}{2}$, there are no real solutions and the real part of $\zeta(p,q')$ vanishes. Note that the symmetry of $q' \leftrightarrow -q'$ has already been taken into account in the first term of~\eqref{eq:zeta}. The $\text{P}$ in the imaginary part of~\eqref{eq:zeta} indicates that the Cauchy principal value is meant to be taken with respect to the $p$ or $q'$ integrals when the system contains a resonance (that is, in the dissipative coupling regime).
\subsection{Dissipative term}
We can now write $\mathbb{L}\hat{\rho}_{\rm S}$ as $\mathbb{L}_{\rm diss}\hat{\rho}_{\rm S}+ \mathbb{L}_{\rm disp}\hat{\rho}_{\rm S}$, and focus in this subsection on the dissipative part. Clearly, in the dispersive coupling regime $\mathbb{L}_{\rm disp}\hat{\rho}_{\rm S} = 0$, as $\left[f(p,q')\right]^{-1}=0$ has no real solutions.

In the dissipative regime where $\left[f(p,q')\right]^{-1}=0$ has elliptical solutions in the $p-q'$ plane,
\begin{align}\label{eq:diss_me}
    \mathbb{L}_{\rm diss}& \hat{\rho}_{\rm S}(t) =  \frac{\pi}{4} \iint_{-\infty}^{\infty} dp dq \left[q_0(p)\right]^{-1}\nonumber \\ & \times \left[\hat{\phi}_{\frac{p}{2} + q_0(p)}\hat{\phi}_{\frac{p}{2}-q_0(p)}\hat{\rho}_{\rm S}(t), \hat{\phi}^{\dagger}_{\frac{p}{2}+q}\hat{\phi}^{\dagger}_{\frac{p}{2}-q}\right] + \text{H.c.}
\end{align}
While this is not a dissipative term in strict Lindblad form, it still gives rise to the dissipative process of photons being emitted from the FH modes with momentum around $\pm q_0(p)$ into the SH. To see this, note that in the interaction picture, the term in~\eqref{eq:diss_me} will oscillate as $\sim \text{exp}\left[i\theta(q^2-q_0^2(p))t\right]$, which averages out to zero except for $q \approx \pm q_0(p)$. Inserting $q = \pm q_0(p)$ exactly into~\eqref{eq:diss_me} would give rise to a Lindblad term with collapse operator $\sim \hat{\phi}_{\frac{s}{2}+q_0(p)}\hat{\phi}_{\frac{s}{2}-q_0(p)}$, but this term would of course be divergent due to the remaining integral over $q$; the non-Lindblad form of Eq.~\eqref{eq:diss_me} ensures a finite decay process by correctly accounting for quantum interference between the continuum of decay channels for modes with momenta $q$ near $\pm q_0(p)$. Note that if we assume the bandwidth of initial excitations to remain limited to sufficiently small momenta, we can replace $q_0(p)$ with $\sqrt{|\xi|}$ in~\eqref{eq:diss_me}, neglecting terms $\mathcal{O}(|\xi|^{-\frac{3}{2}})$.

In the spatial domain, we have
\begin{align}
    \mathbb{L}_{\rm diss}&\hat{\rho}_{\rm S}(t) = \frac{\pi}{4\sqrt{|\xi|}} \iint_{-\infty}^{\infty} dy dy' e^{2\pi i \sqrt{|\xi|}(y-y')}\nonumber \\ & \times \left[\hat{\Phi}_y\hat{\Phi}_{y'}\hat{\rho}_{\rm S}(t),\hat{\Phi}^{\dagger 2}_{(y+y')/2}\right] + \text{H.c.} + \mathcal{O}(|\xi|^{-\frac{3}{2}}),
\end{align}
which describes correlated two-photon emission processes between spatial fields between $y$ and $y'$ with correlation length $|\xi|^{-\frac{1}{2}}$. Clearly as $|\xi| \rightarrow \infty$, this converges to a spatially local two-photon loss.

\subsection{Dispersive term}


The dispersive term is 
\begin{align}\label{eq:disp_me}
    \mathbb{L}_{\rm diss}& \hat{\rho}_{\rm S}(t) = -\frac{i}{2}\text{P}\iiint_{-\infty}^{\infty} \!  dp dq dq' f(p,q') \nonumber \\ & \times \left[\hat{\phi}_{\frac{p}{2} + q'}\hat{\phi}_{\frac{p}{2}-q'}\hat{\rho}_{\rm S}, \hat{\phi}^{\dagger}_{\frac{p}{2}+q}\hat{\phi}^{\dagger}_{\frac{p}{2}-q}\right] + \text{H.c.}
\end{align}

To inspect the role that this term plays on the reduced FH dynamics, we consider dissipative and dispersive coupling regimes separately.
\subsubsection{Dispersive coupling regime}
First, we consider the dispersive coupling regime where $\xi \gg 1$, and $\beta < 1/2$, such that $\left[f(q,p')\right]^{-1}$ has no zeroes. We can thus drop the Cauchy principal value. In this regime, provided the bandwidth of initial excitations is not too large, we can approximate $f(p,q') \approx -\frac{\theta}{2\xi} + \mathcal{O}(\xi^{-2})$, as we assume that states with significant momenta $\frac{p}{2} \pm q$ remain unpopulated. Then, we can find $\mathbb{L}_{\rm disp}\hat{\rho}_{\rm S} = -i[\hat{W}_{\rm SPM},\hat{\rho}_{\rm S}]$, where
\begin{equation}
\hat{W}_{\rm SPM} = \frac{\theta}{4\xi}\iiint^\infty_{-\infty}dpdqdq'\hat{\phi}_{\frac{p}{2}+q}^\dagger\hat{\phi}_{\frac{p}{2}-q}^\dagger\hat{\phi}_{\frac{p}{2}+q'}\hat{\phi}_{\frac{p}{2}-q'},
\end{equation}
which is precisely that obtained by the leading order SW transformation result in Sec.~\ref{sec:dispersive_coupling}. Note that since the unitary transformation only provides corrections of order $\sim |\xi|^{-2}$ and higher, the ME and SW formalism both result in the same conventional multimode cubic cascaded nonlinearity Hamiltonian, and are thus in complete qualitative agreement in the dispersive coupling regime, and also agree quantitatively to leading order.

In the spatial basis, the SPM Hamiltonian becomes
\begin{equation}
    \hat{W}_{\rm SPM} = \frac{\theta}{4\xi}\int dy \hat{\Phi}^{\dagger 2}_y \hat{\Phi}^2_y.
\end{equation}
as expected.

\subsubsection{Dissipative coupling regime}

In the dissipative coupling regime, we must take more care as $\left[f(p,q')\right]^{-1}=0$ now has real solutions, and the Cauchy principal value must be paid attention in~\eqref{eq:disp_me}. Here we focus again on the case with elliptical solutions, which is defined by (provided $\gamma$ is sufficiently small) $\xi \ll -1$ and $\beta < 1/2$.

It is convenient here to again use the intraband and extraband notation from Sec.~\ref{sec:dissipative_coupling}. Within the intraband regime, we can again expand $f(p,q') \approx -\frac{\theta}{2\xi} + \mathcal{O}(\xi^{-2})$, such that, similar to the dispersive coupling case, we can drop the Cauchy principal value and find
\begin{align}
\hat{W}_{\rm i, SPM} = \frac{\theta}{4\xi}&\int_{-2p_{\rm i}}^{2p_{\rm i}} dp\iint^{q_{\rm i}(p)}_{-q_{\rm i}(p)}dqdq'dp \nonumber \\ & \times \hat{\phi}_{\frac{p}{2}+q}^\dagger\hat{\phi}_{\frac{p}{2}-q}^\dagger\hat{\phi}_{\frac{p}{2}+q'}\hat{\phi}_{\frac{p}{2}-q'},
\end{align}
again in agreement with the leading-order results derived in Sec.~\ref{sec:dissipative_coupling} using the SW transformation (and ME for the extraband component as the reservoir).

In extraband regime, $f(p,q')$ becomes very sharply peaked around $q' = \pm q_0(p)$. Thus, to take the Cauchy principal value, one can assume that in a very narrow band around $q' = \pm q_0(p)$ the FH operators in Eq.~\eqref{eq:disp_me} can be approximated as $\hat{\phi}_{\frac{s}{2} \pm q_0(p)}$, and the Cauchy principal value integral over $q'$ vanishes as $\text{P}\int dx/(x^2-|a|) =0$ for any $a$. Following the same logic, for the extraband regime, we can expect the dispersive term to contribute only a small nonlinear Lamb shift-like dispersive component to the Hamiltonian, and it should thus suffice to retain only the Hamiltonian $\hat{W}_{\rm i, SPM}$ term, to leading order, thus recovering exactly the leading order results from the SW transformation (neglecting the higher order processes). In any case, since photons in the extraband regime in the dissipative coupling regime will exponentially decay, the system should ultimately localize itself exclusively in the intraband regime.

\section{Derivation of ME for dissipative coupling regime and regime of validity for Markov approximation}\label{app:diss_me}
\subsection{Derivation of ME}

In this subsection, we derive a ME for the system intra-band component of the FH modes, by treating the extra-band component as a reservoir. This derivation follows very closely that of Appendix.~\ref{app:master_equation}, so we only sketch the key points.

The total Hamiltonian is $\hat{H} = \hat{H}_{\rm L} + \hat{V}_{\rm i} +\hat{V}_{\rm e}$, as defined in Sec.~\ref{sec:dissipative_coupling}. To derive a ME, we will partition the Hilbert space of the system such that the ``extra-band'' component of the FH acts as a reservoir field, and the ``system'' consists of the SH field and the intra-band FH modes. In this manner, we can split up the Hamiltonian into system, interaction, and reservoir parts as $\hat{H} = \hat{H}_{\rm S}+ \hat{H}_{\rm int} + \hat{H}_{\rm R}$, where $\hat{H}_{\rm S} = \hat{H}_{\rm L}^{\rm S} + \hat{V}_{\rm i}$, where
\begin{equation}
\hat{H}_{\rm L}^{\rm S} = \theta  \left[ \int_{-p_{\rm i}}^{p_{\rm i}} \! \! dp\frac{p^2}{2} \hat{\phi}_p^{\dagger}\hat{\phi}_p \!\!+\!\! \int_{-\infty}^{\infty}dp(-\xi + \gamma s +\frac{\beta p^2}{2})\hat{\psi}^{\dagger}_p\hat{\psi}_p\right],
\end{equation}
\begin{equation}
\hat{V}_{\rm i} = \frac{1}{2}\int_{-2p_{\rm i}}^{2p_{\rm i}} dp \int_{-q_{\rm i}(p)}^{q_{\rm i}(p)}\!\! dq \hat{\psi}^{\dagger}_p\hat{\phi}_{\frac{p}{2} + q}\hat{\phi}_{\frac{p}{2} - q} + \text{H.c.},
\end{equation}
and  $\hat{H}_{\rm int} = \hat{V}_e$:
\begin{equation}
\hat{V}_{\rm e} = \int_{-\infty}^{\infty} dp \int_{q_{\rm i}(p)}
^{\infty} \! \! dq \hat{\psi}^{\dagger}_p \hat{R}_{\frac{p}{2} + q}\hat{R}_{\frac{p}{2} - q}+ \text{H.c.},
\end{equation}
where we let $q_{\rm i}(p) =0$ for $|p| > 2p_{\rm i}$, and we let $\hat{\phi}_p = \hat{R}_p$ for $|p| > p_{\rm i}$ to clarify that it is a reservoir operator. Finally, the reservoir Hamiltonian is $\hat{H}_{\rm R} = \hat{H}_{\rm L}^{\rm R}$:
\begin{equation}
\hat{H}_{\rm L}^{\rm R} = \int_{p_{\rm i}}^{\infty} dp \frac{p^2}{2}\left( \hat{R}_p^{\dagger} \hat{R}_p + \hat{R}_{-p}^{\dagger} \hat{R}_{-p}\right).
\end{equation}
To derive a Born-Markov ME for the reduced system dynamics, we then make a second-order Born-Markov approximation and ``second Markov approximation'', and assume an initial vacuum state $\hat{\rho}_{\rm R}$ for the reservoir modes. This results in a system density matrix equation of motion
 $\frac{ d}{dt}\hat{\rho}_{\rm S}(t) = -i[\hat{H}_{\rm S},\hat{\rho}_{\rm S}(t)] + \mathbb{L}\hat{\rho}_{\rm S}(t)$, where 
\begin{equation}
\mathbb{L}\hat{\rho}_{\rm S}= \int_0^{\infty} d\tau \text{Tr}_{\rm R}\Big[\hat{\tilde{V}}_{\rm e}(-\tau)\hat{\rho}_{\rm S}\hat{\rho}_{\rm R},\hat{V}_{\rm e}\Big] + \text{H.c.},
\end{equation}
and $\hat{\tilde{V}}_{\rm e}(-\tau)$ denotes the interaction picture at time $-\tau$. We assume that the interaction picture can be 
moved into by neglecting $\hat{V}_{\rm i}$, since $\hat{H}_{\rm S} = \hat{H}_{\rm L}^{\rm S} + \mathcal{O}(|\xi|^0)$, and including $\hat{V}_{\rm i}$ would only provide sub-leading order corrections. Thus, we have
\begin{equation}
\hat{\tilde{V}}_{\rm e}(t) = \int_{-\infty}^{\infty} dp \int_{q_{\rm i}(p)}
^{\infty} \! \! dq \hat{\psi}^{\dagger}_p \hat{R}_{\frac{p}{2} + q}\hat{R}_{\frac{p}{2} - q} e^{\frac{i}{2}\left[f(p,q)\right]^{-1}t}+ \text{H.c.}.
\end{equation}
 Performing the partial trace and evaluating vacuum expectation values, we find,
\begin{equation}\label{eq:7}
\mathbb{L}\hat{\rho}_{\rm S} =  \int_{-\infty}^{\infty} \!\!\! dp \int_{q_{\rm i}(p)}^\infty \!\! dq[\hat{\psi}_p\hat{\rho}_{\rm S},\hat{\psi}_p^{\dagger}] \int_0^{\infty}\!\!\! d\tau e^{\frac{i}{2}\left[f(p,q)\right]^{-1}\tau} + \text{H.c.}
\end{equation}
We then evaluate the $\tau$-integral by using the relation $\int_0^{\infty}d\tau e^{i\omega \tau} = \pi\delta(\omega) + \text{P}\frac{i}{\omega}$, where $P$ is the Cauchy principal value. By definition of the dissipative coupling regime, we assume $q_0(p) \approx \sqrt{|\xi|}$ is greater than $q_{\rm i}(p)$ for all $p$. 

The result is that we can describe the influence of the extra-band interaction by adding to the system Hamiltonian a Lamb-shift type term given by Eq.~\eqref{eq:G_ls}, in terms of the frequency shift of the SH given by 
\begin{align}
    \delta(p) = -\frac{\theta}{q_0(p)}\text{arctanh}\left(\frac{q_{\rm i}(p)}{q_0(p)}\right) = -\theta\frac{q_{\rm i}(p)}{|\xi|} + \mathcal{O}(|\xi|^{-\frac{3}{2}}).
\end{align}Additionally, there is a continuous Lindblad term $\hat{L}_p = \sqrt{\kappa(p)}\hat{\psi}_p$, where $\kappa(p)$ is given by 
\begin{equation}
\kappa(p) = \frac{\pi}{q_0(p)} = \frac{\pi}{\sqrt{|\xi|}} + \mathcal{O}(|\xi|^{-\frac{3}{2}}).
\end{equation}

\subsection{Validity of Born-Markov approximation}

In this section we examine in some qualitative detail the validity of the second-order Born-Markov approximation. For simplicity we focus on the derivation in Appendix~\ref{app:diss_me}, where we treat the extraband component as the reservoir coupled to the SH. In the case where the SH is the reservoir, which we treat in Appendix.~\ref{app:master_equation}, the analysis is made more complicated by the fact that the memory kernel of the reservoir is itself tied to the dynamics of the system, due to the integration over a continuous mode index. Nonetheless, we expect the general argument outlined in this section to also be broadly applicable to Appendix.~\ref{app:master_equation}.

To investigate the validity of the Born-Markov approximation, one needs to compare the memory timescale of the reservoir to the coupling rates of the system with the reservoir. One way to do so is by inspecting the Nakajima-Zwanzig~\cite{breuer2002} form of the ME:
\begin{equation}\label{eq:NZ}
\frac{ d}{dt}\hat{\tilde{\rho}}_{\rm S}(t)= \int_0^{\infty} d\tau \text{Tr}_{\rm R}\Big[\hat{\tilde{V}}_{\rm e}(t-\tau)\hat{\tilde{\rho}}_{\rm S}(t-\tau)\hat{\rho}_{\rm R},\hat{\tilde{V}}_{\rm e}(t)\Big] + \text{H.c.},
\end{equation}
where we have moved back into the interaction picture. The Nakajima-Zwanzig form of the ME is expected to be accurate to the same level of perturbation theory as the standard Born-Markov approximation. We can use this form to investigate the quantitative conditions under which the ``Markov approximation'' can be made---that is, replacing $\hat{\tilde{\rho}}_{\rm S}(t-\tau)$ with $\hat{\tilde{\rho}}(t)$, which gives us the Born-Markov ME derived in Appendix~\ref{app:diss_me}. Since the Born approximation, which gives the Nakajima-Zwanzig form in~\eqref{eq:NZ}, is valid under the same conditions as the Markov approximation~\cite{breuer2002}, we can use this as a test of the regime of validity of the Born-Markov ME as a whole.

Now, using the Nakajima-Zwanzig form, we proceed to the same level of the derivation as in Eq.~\eqref{eq:7}, but this time we evaluate the $q$ integral first:

\begin{equation}
\frac{ d}{dt}\hat{\tilde{\rho}}_{\rm S}(t)= \int_0^{\infty} \!\!\!d\tau \int_{-\infty}^{\infty} \!\!\! dp \mathcal{K}_p(\tau)\left[\hat{\psi}_p\hat{\tilde{\rho}}_{\rm S}(t\!-\!\tau),\hat{\psi}^{\dagger}_p\right] + \text{H.c.},
\end{equation}
where $\mathcal{K}_p(\tau)$ is the memory kernel of the system-reservoir interaction:
\begin{align}
\mathcal{K}_p(\tau) & = \int_{q_{\rm i}(p)}^{\infty} dq e^{i\theta \left[q^2 - q_0^2(p)\right]\tau} \nonumber \\
& = \sqrt{\frac{\pi}{4\tau}}e^{i\theta\left[\frac{\pi}{4} - q_0^2(p)\tau\right]}\left[1-\text{erf}\left(e^{-i\theta\frac{\pi}{4}}q_{\rm i}(p)\sqrt{\tau}\right)\right].
\end{align}

Now, from the ME derived in Appendix~\ref{app:diss_me}, the leading order term in the perturbative parameter $|\xi|^{-1}$ is the decay term which scales as $\kappa \sim |\xi|^{-\frac{1}{2}}$. $\hat{\rho}_{\rm S}(t-\tau)$. Thus, we can consider the characteristic timescale of $\hat{\tilde{\rho}}_{\rm S}(t-\tau)$ to be $|\xi|^{\frac{1}{2}}$. We can thus compute the value of the memory kernel at this value of the time delay integration parameter $\tau$:
\begin{equation}
\mathcal{K}_p(|\xi|^{\frac{1}{2}})\approx\begin{cases} 
      i\frac{\theta}{2}\sqrt{\frac{\pi}{2}}e^{-i\theta |\xi|^{\frac{3}{2}}}|\xi|^{-\frac{1}{2}} & |p| \ll 2p_{\rm i} \\
      \sqrt{\frac{\pi}{4}}e^{i\theta\left[\frac{\pi}{4} - q_0^2(p)|\xi|^{\frac{1}{2}}\right]}|\xi|^{-\frac{1}{4}} & |p| > 2p_{\rm i}
   \end{cases}
\end{equation}
Over the characteristic timescale of evolution of the interaction reduced density operator, the memory kernel either decays as $\sim |\xi|^{1/2}$ with an accumulated phase $\sim |\xi|^{3/2}$ in the small SH momentum $p$ case, or decay as  $\sim |\xi|^{1/4}$ with an accumulated phase $\sim |\xi|^{1/2}$ in the large $p$ case. In either case, for $|\xi| \gg 1$, the contributions from $\hat{\rho}_{\rm S}(t-\tau) - \hat{\rho}_{\rm S}(t)$ average out (and decay in amplitude slowly) to zero. Thus the Markov approximation is valid under the limit $|\xi| \rightarrow \infty$, as expected.

A more rigorous and quantitative estimate of the error associated with the Born-Markov approximation would require a deeper analysis, as standard with ME treatments.
Since the interaction Hamiltonian is essentially the same in the case where the SH is treated as the reservoir for the FW, we expect some of the qualitative arguments we put forth in this appendix to also apply to that case.

\section{Single mode limit}\label{app:single}

In this appendix we consider a ``single-mode'' limit of the cascaded $\chi^{(2)}$ model described Sec.~\ref{sec:dissipative_coupling} in the dissipative coupling regime. It is important to note that the dissipative nature of this regime is an intrinsically multi-mode effect; the purpose of this section is merely to demonstrate concepts from the main text with a simpler toy model which retains the key features, while being more intuitive and allowing for easier numerical simulations.

In the single-mode limit, we consider the Hamiltonian $\hat{H} = \hat{H}_0 + \hat{V}$, where $\hat{H}_0 = \xi \hat{\psi}^{\dagger} \hat{\psi}$, $\hat{V} = \frac{1}{2}\hat{\psi}^{\dagger}\hat{\phi}^2 + \text{H.c.}$, and $[\hat{\psi},\hat{\psi}^{\dagger}] = [\hat{\phi},\hat{\phi}^{\dagger}]=1$. We also consider the Lindblad operator $\hat{L} = \sqrt{\pi} \xi^{-\frac{1}{4}}\hat{\psi}$; note that we can consider $\xi \gg 1$ for the purposes of this section. By considering the unitary transformation $e^{\hat{S}}$ with $\hat{S} = [2\xi]^{-1}(\hat{\psi}^{\dagger}\hat{\phi}^2-\text{H.c.})$, we similarly find $\hat{H}' = \hat{H}_0 + \hat{W}_{\rm lin} + \hat{W}_{\rm SPM} + \hat{W}_{\rm XPM}$, with $\hat{W}_{\rm lin} = \frac{1}{2\xi}\hat{\psi}^{\dagger}\hat{\psi}$, $\hat{W}_{\rm SPM} = -\frac{1}{4\xi}\hat{\phi}^{\dagger 2}\hat{\phi}^2$, and $\hat{W}_{\rm XPM} = \frac{1}{\xi}\hat{\psi}^{\dagger}\hat{\psi}\hat{\phi}^{\dagger}\hat{\phi}$. The Lindblad transforms as $\hat{L}' = -\sqrt{\pi/2}\xi^{-\frac{5}{4}}\hat{\phi}^2$.

\begin{figure}
\centering 
\includegraphics[width=0.48\textwidth]{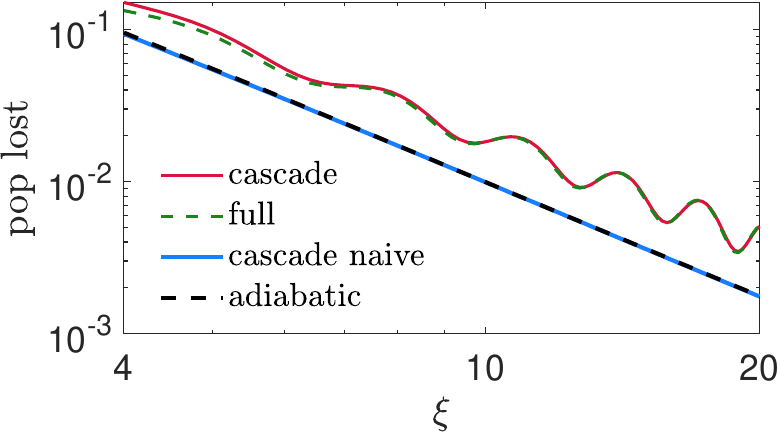}
    \caption{Population loss $(N_{\rm FW}(t=0)-N_{\rm FW} )/N_{\rm FW}(t=0)$ from an initial lab-frame two-photon Fock state. The full lab-frame Hamiltonian simulation is shown in green, and the SW-transformation cascaded nonlinearity Hamiltonian $\hat{H}'$ and Lindblad $\hat{L'}$ are shown in solid red. The blue lines show the cascade solution \emph{without accounting} for the SW transformation in either the observable~\eqref{eq:E1} or the initial state. The dashed black line corresponds to an adiabatic preparation of the dressed FH state with $\xi_i = 200$ and $T=9$, using a linear $\xi(t)$ function with ramp duration $T$.
    }
    \label{fig:appfig}
\end{figure} 

Again, using the mean field approximation we can calculate expectation values. For instance, photon number is 
\begin{equation}\label{eq:E1}
    N_{\rm FW} = \langle \hat{\phi}^{\dagger }\hat{\phi} \rangle + \frac{1}{\xi}\left[\langle \hat{\psi} \hat{\phi}^{\dagger 2}\rangle + \text{H.c.}\right] -\frac{1}{2\xi^2}\langle \hat{\phi}^{2\dagger} \hat{\phi}^2\rangle,
\end{equation}
and the initial condition in the dressed frame can be expressed as $ e^{\hat{S}}\ket{\phi_0}$,
where $\ket{\phi_0}$ is the lab frame initial state, which is assumed to be the vacuum state for the SH.

As with in Sec.~\ref{sec:mf}, we can calculate the term containing initial entanglement as 
\begin{align}
    \langle \hat{\psi} \hat{\phi}^{\dagger 2}\rangle &= \text{Tr}\left[\hat{\phi}^{\dagger 2} e^{\mathbb{L}t}\hat{\psi}\hat{S}\ket{\phi_0}\bra{\phi_0})\right]e^{-i\xi t - \frac{\pi}{{2\sqrt{\xi}}}t} \\
    & = \frac{1}{2\xi}\langle \hat{\phi}^{\dagger 2}\hat{\phi}^2 \rangle(t=0) e^{-i\xi t - \frac{\pi}{2\sqrt{\xi}}t},
\end{align}
where in the second line we have neglected the higher-order effects of self-phase modulation and two-photon loss in the FW.

Similarly to Sec.~\ref{sec:mf}, we can excite the dressed FH modes in the SW frame without populating the dressed SH excitations by adiabatically varying the phase-mismatch $\xi \rightarrow \xi(t)$,
where the initial phase-mismatch $\xi_i$ is much larger than the final phase-mismatch $\xi_f$.

In Fig.~\ref{fig:appfig}, we plot the population lost for an initial FH coherent state in the lab-frame. As expected, one sees quantitative agreement \emph{only} with the higher-order corrections from ~\eqref{eq:E1} and using the dressed-frame initial condition. The naive solution which simply uses the cascade Hamiltonian $\hat{H} = \hat{W}_{\rm SPM}$ and two-photon Lindblad $\hat{L'}$ without accounting for the SW transformation, however, faithfully represents the case where the phase-mismatch is adiabatically varied to prepare the system in the dressed FH state, confirming the efficacy of our proposal in Sec.~\ref{sec:adiabatic}.

\end{appendix}
\bibliography{arxiv_manuscript.bib}
\end{document}